\documentclass[12pt,a4paper,reqno]{amsart}
\pdfoutput=1

\usepackage{amsmath,amssymb,a4}
\usepackage[breaklinks=true]{hyperref}
\usepackage[strings]{underscore}
\usepackage{enumerate}

\newtheorem{lemma}{Lemma}
\newtheorem{corollary}[lemma]{Corollary}

\newtheorem{ansatz}[lemma]{Ansatz}
\newtheorem{proposition}[lemma]{Proposition}
\newtheorem{theorem}[lemma]{Theorem}
\theoremstyle{remark}
\newtheorem{remark}[lemma]{Remark}
\DeclareMathOperator{\Res}{Res}

\allowdisplaybreaks[4]

\begin{document}

\title[An algebraic approach to a quartic 
Kontsevich model]{An algebraic approach to a quartic analogue of the 
Kontsevich model}

\author{J\"org Sch\"urmann}
\author{Raimar Wulkenhaar}

\address{Mathematisches Institut der WWU, 
Einsteinstr.\ 62, 48149 M\"unster, Germany\\
{\itshape E-mail adresses:} \normalfont 
\texttt{jschuerm@uni-muenster.de}, \texttt{raimar@math.uni-muenster.de}}

\begin{abstract}
  We consider an analogue of Kontsevich's matrix Airy function where
  the cubic potential $\mathrm{Tr}(\Phi^3)$ is replaced by a quartic
  term $\mathrm{Tr}(\Phi^4)$. Cumulants of the resulting measure are
  known to decompose into cycle types for which a recursive system of
  equations can be established. We develop a new, purely algebraic
  geometrical solution strategy for the two initial equations of the
  recursion, based on properties of Cauchy matrices. These structures
  led in subsequent work to the discovery that
  the quartic analogue of the Kontsevich model obeys blobbed topological
  recursion.
\end{abstract}


\maketitle
\markboth{\hfill\textsc\shortauthors}{\textsc{{An algebraic
      approach to a quartic analogue of the Kontsevich model}\hfill}}

\section{Introduction}

Guided by uniqueness of quantum gravity in two dimensions, Witten
conjectured in \cite{Witten:1990hr} that the generating function of
intersection numbers of tautological characteristic classes on the
moduli space of stable complex curves has to satisfy the PDE of the
Korteweg-de Vries hierarchy. The conjecture was proved a few months
later by Kontsevich in his seminal paper
\cite{Kontsevich:1992ti}. Kontsevich understood that critical graphs
of the canonical Strebel differential \cite{Strebel} on a punctured
curve give a cell-decomposition of the moduli space of punctured
curves, which can be organised into a novel type of matrix model (the
`matrix Airy function') with covariance
\[
\langle \Phi(e_{jk})\Phi(e_{lm})\rangle_c=
\frac{\delta_{kl}\delta_{jm}}{ \lambda_j+\lambda_k}
\]
(where $(e_{jk})$ denotes the standard matrix basis and $\delta_{kl}$ 
the Kronecker symbol) and tri-valent
vertices. The $\lambda_j$ are Laplace transform
parameters\footnote{These $\lambda_j$ will be denoted by $E_j$ in this
paper.} of the
lengths $L_j$ of critical trajectories of the Strebel
differentials, and the 
generically simple zeros of the Strebel differential correspond to 
tri-valent vertices. Kontsevich went on to establish that the
logarithm of the partition function of his matrix model is the
$\tau$-function for the KdV-hierarchy, thereby proving that 
his matrix model is integrable.

The same covariance (up to normalisation)
\[
\langle \Phi(e_{jk})\Phi(e_{lm})\rangle_c=
\frac{\delta_{kl}\delta_{jm}}{ N(E_j+E_k)}
\] 
arises in quantum field theory models on noncommutative geometries
\cite{Grosse:2005ig}, where the $E_k$ are the spectral values (`energy
levels') of a Laplace-type operator. These are models for scalar fields
with cubic self-interaction. From a quantum field theoretical point of
view one would be more interested in a quartic self-interaction, which
e.g.\ is characteristic to the Higgs field. Such quartic models have
been understood in \cite{Grosse:2004yu} at the level of formal power
series. Later in \cite{Grosse:2009pa, Grosse:2012uv} exact equations
between correlation functions in the quartic (matrix) model were
derived. These equations share many aspects with a universal structure
called topological recursion \cite{Eynard:2007kz}.

Such recursions typically rely on the initial solution of a non-linear
problem (for the Kontsevich model achieved in \cite{Makeenko:1991ec}).
For the quartic model, the corresponding equation (for the planar
two-point function of cycle type $(0,1)$) is given in (\ref{eq:GZW})
below. Its solution succeeded in \cite{Grosse:2019jnv}, via a larger
detour. It was assumed that (\ref{eq:GZW}) converges for $N\to \infty$
to an integral equation with H\"older-continuous measure. The special
case of constant measure was solved in \cite{Panzer:2018tvy} with help
from computer algebra. Its structure suggested a conjecture for the
general case which was proved in \cite{Grosse:2019jnv} by residue
theorem and Lagrange-B\"urmann resummation.

This paper provides a novel algebraic geometrical solution strategy
for the non-linear equation (\ref{eq:GZW}) and the affine equation
(\ref{calG11:GW}) (which determines the planar two-point function of
cycle type $(2,0)$). We (re)prove that these cumulants are
compositions of rational functions with a preferred inverse of another
rational function
\[
  R(z)=z-\frac{\lambda}{N} \sum_{k=1}^d
  \frac{\varrho_k}{z+\varepsilon_k}\:.
\]

Building on these results it was understood in \cite{Branahl:2020yru} that
derivatives of the partially summed two-point function with respect to the
spectral values $E_k$ extend to meromorphic differentials
$\omega_{g,n}$ labelled by genus $g$ and number $n$ of marked points
of a complex curve. The $\omega_{g,n}$ are supplemented by
two families of auxiliary functions and satisfy a coupled
system of equations.  The solution of this system 
for small $-\chi=2g+n-2$ in \cite{Branahl:2020yru} gave strong support
for the conjecture that the $\omega_{g,n}$ obey blobbed
topological recursion \cite{Borot:2015hna} for the spectral curve
$(x:\hat{\mathbb{C}}\to\hat{\mathbb{C}}, \omega_{0,1}=xdy,\omega_{0,2})$
given by
\[
  x(z)=R(z)\;,\qquad
  y(z)=-R(-z)\;,\qquad
  \omega_{0,2}(w,z)=\frac{dw\,dz}{(w-z)^2}+\frac{dw\,dz}{(w+z)^2}\;.
\]
The proof of this conjecture for $g=0$ was
achieved in \cite{Hock:2021tbl}. As shown in \cite{Borot:2015hna},
blobbed topological recursion generates intersection numbers
on the moduli space $\overline{\mathcal{M}}_{g,n}$ of stable
complex curves. In view of
the deep r\^ole played by the global involution
$z\mapsto -z$ \cite{Hock:2021tbl} we expect that this
very natural involution will find a counterpart in
the intersection theory encoded in the quartic analogue of the
Kontsevich model. Working out the details is a fascinating programme
left for the future.

\section*{Acknowledgements}\enlargethispage{2.6mm}

Our work was supported\footnote{``Gef\"ordert durch die Deutsche
  Forschungsgemeinschaft (DFG) im Rahmen der Exzellenz\-strategie des
  Bundes und der L\"ander EXC 2044 -- 390685587, Mathematik M\"unster:
  Dynamik--Geometrie--Struktur"} by the Cluster of 
Excellence \emph{Mathematics M\"unster}. RW would like to thank Harald
Grosse and Alexander Hock for the collaboration which provided the
basis for the present paper.

\section{Matrix integrals}

\label{sec:matrixintegrals}

Let $H_{N}$ be the real vector space of self-adjoint $N\times
N$-matrices and $(E_1,\dots, E_{N})$ be not necessarily distinct positive
real numbers. By the Bochner-Minlos theorem \cite{Bochner}, combined
with the Schur product theorem \cite[\S 4]{Schur}, there is a unique
probability measure $d\mu_0(\Phi)$ on the dual space $H_{N}'$ with
\begin{align}
\exp\Big(- \frac{1}{2N} \sum_{k,l=1}^{N}
\frac{M_{kl}M_{lk}}{E_k+E_l}\Big)
= \int_{H_{N}'} d\mu_0(\Phi) \,e^{\mathrm{i} \Phi(M)}
\label{measure0}
\end{align}
for any $M=M^*=\sum_{k,l=1}^{N} M_{kl} e_{kl}\in H_{N}$, 
where $(e_{kl})$ is
the standard matrix basis. The linear forms extend via 
$\Phi(M_1+\mathrm{i}M_2):=\Phi(M_1)+\mathrm{i}\Phi(M_2)$ to arbitrary complex 
$N\times N$-matrices. This allows us to
evaluate $\Phi(e_{jk})$ and to identify the covariance 
\[
\int_{H_{N}'} d\mu_0(\Phi) \,
\Phi(e_{jk})\Phi(e_{lm}) = \frac{\delta_{kl} \delta_{jm}}{N (E_j+E_k)}\;.
\]

We are going to deform the Gau\ss{}ian 
measure (\ref{measure0}) by a quartic potential, 
\begin{align}
d\mu_\lambda(\Phi) &:= 
\frac{d\mu_0(\Phi)\; \mathcal{P}_4(\Phi,\lambda)}{
\int_{H_{N}'} d\mu_0(\Phi)\;\mathcal{P}_4(\Phi,\lambda)}\;,\qquad
\label{measure4}
\\
\mathcal{P}_4(\Phi,\lambda)&= 
\exp\Big({-}\,\frac{\lambda N}{4} \mathrm{Tr}(\Phi^4)\Big):=
\exp\Big({-}\,\frac{\lambda N}{4} \!\!
\sum_{j,k,l,m=1}^{N} \!\!\!\!\! \Phi(e_{jk})\Phi(e_{kl})\Phi(e_{lm})\Phi(e_{mj})\Big)\;,
\nonumber
\end{align}
for some $\lambda>0$. 
This matrix measure is the quartic analogue of
the Kontsevich model \cite{Kontsevich:1992ti} in which the
deformation is given by the cubic term
\[
\mathcal{P}_3(\Phi,\lambda)
=\exp\Big({-}\,\frac{\lambda N}{3} \mathrm{Tr}(\Phi^3)\Big)
:=\exp\Big(-\frac{\lambda N}{3} 
\sum_{k,l,m=1}^{N} \Phi(e_{kl})\Phi(e_{lm})\Phi(e_{mk})
\Big)\:.
\] 
The cubic measure was designed to prove Witten's
conjecture \cite{Witten:1990hr} that intersection
numbers of tautological characteristic classes on the moduli space of
stable complex curves are related to the KdV hierarchy. Kontsevich
proved that $\log \int_{H'_N} d\mu_0(\Phi)\;
\mathcal{P}_3(\Phi,\frac{\mathrm{i}}{2})$, viewed as function of
$t_k=-(2k-1)!!\frac{1}{N}\sum_{j=1}^{N} E_j^{-(2k+1)}$,
is the generating function of these 
intersection numbers.

\bigskip

We are interested in moments of the measure (\ref{measure4}),
\begin{align}
\langle e_{k_1l_1}\dots e_{k_nl_n}\rangle &:= \int_{H'_{N}} \!\! 
d\mu_\lambda(\Phi)
\;\Phi(e_{k_1l_1}) \cdots \Phi(e_{k_nl_n}) 
= \frac{1}{\mathrm{i}^{n}}
\frac{\partial^n\mathcal{Z}(M)}{\partial 
M_{k_1l_1}\cdots \partial M_{k_nl_n}} \Big|_{M=0}\;,
\nonumber
\\
\mathcal{Z}(M)&= \int_{H'_N}\!\! d\mu_\lambda(\Phi) \;e^{\mathrm{i}\Phi(M)}
\;.
\label{moments}
\end{align}
As explained in Appendix~\ref{appA} (see also
\cite{McCullagh2012,Speed}), the moments (\ref{moments}) decompose
into cumulants
\begin{align}
\Big\langle \prod_{i=1}^n e_{k_il_i}\Big\rangle 
=\sum_{\substack{\text{partitions} \\ \text{$\pi$ of 
$\{1,\dots, n\}$}}} \prod_{\text{blocks $\beta \in \pi$}} 
\Big\langle \prod_{i\in \beta}  e_{k_i l_i} \Big\rangle_c\;.
\end{align}
For a quartic potential (\ref{measure4}), moments and cumulants are
only non-zero if $n$ is even and every block $\beta$ is of even
length.  The structure of the Gau\ss{}ian measure (\ref{measure0})
(together with the invariance of a trace under cyclic permutations)
implies that $\langle e_{k_1l_1}\dots e_{k_nl_n}\rangle_c$ is only
non-zero if $(l_1,\dots,l_n)=(k_{\sigma(1)},\dots,k_{\sigma(n)})$ is a
permutation of $(k_1,\dots,k_n)$, and in this case the cumulant only
depends on the \emph{cycle type} of this permutation $\sigma$ in the
symmetric group $ \mathcal{S}_n$ (see Appendix~\ref{appA}, with $b\geq
1$ the number of cycles of length $n_i>0$, $n_1+\cdots +n_b=n$):
\begin{align}
N^{n_1+\dots+n_b}
\big\langle (e_{k_1^1k_2^1} 
e_{k_2^1k_3^1} \cdots 
e_{k_{n_1}^1k_1^1}) \cdots 
(e_{k_1^bk_2^b} e_{k_2^bk_3^b} \cdots 
e_{k_{n_b}^bk_1^b}) \big\rangle_c
\nonumber
\\
=: 
N^{2-b} G_{|k_1^1\dots k_{n_1}^1|\dots
|k_1^b\dots k_{n_b}^b|} \;.
\label{eq:Ggbn-all}
\end{align}
To correctly identify the cycles of the permutation it is necessary
that all $k^i_j$ are pairwise different in (\ref{eq:Ggbn-all}). These
$N$-rescaled cumulants (\ref{eq:Ggbn-all}) are further expanded as
formal power series
$G_{\dots}=\sum_{g=0}^\infty \frac{1}{N^{2g}} G^{(g)}_{\dots} $ in
$N^{-2}$ so that
\begin{align}
N^{n_1+\dots+n_b} \big\langle (e_{k_1^1k_2^1} 
e_{k_2^1k_3^1} \cdots 
e_{k_{n_1}^1k_1^1}) \cdots 
(e_{k_1^bk_2^b} e_{k_2^bk_3^b} \cdots 
e_{k_{n_b}^bk_1^b}) \big\rangle_c \nonumber
\\
= \sum_{g=0}^\infty 
N^{2-b-2g} \cdot G^{(g)}_{|k_1^1\dots k_{n_1}^1|\dots
|k_1^b\dots k_{n_b}^b|} \;.
\label{eq:Ggbn}
\end{align}
It turns out that this grading $(g,b)$ of 
$G^{(g)}_{|k_1^1\dots k_{n_1}^1|\dots |k_1^b\dots k_{n_b}^b|}$
fits with the combinatorics of \emph{ribbon graphs} (with
$4$-valent vertices) on a \emph{connected} oriented compact
topological surface of genus $g\geq 0$ with $b\geq 1$ boundary
components (and $n_i$ labels on the $i^{\text{th}}$ boundary
component) and \emph{Euler characteristic} $\chi=2-2g-b$ (see e.g.\
\cite[\S 3]{Grosse:2003aj} for the particular case of $4$-valent
vertices, and compare also with \cite{Kontsevich:1992ti} or \cite[\S 2
and \S 6]{Eynard:2016yaa}). Note that the moments are related to
ribbon graphs on possibly \emph{non-connected} oriented compact
topological surfaces (see e.g.\ \cite[\S 3, Prop.3.8.3]{Lando}).

Starting point for the investigation of cumulants are equations of motion for 
$\mathcal{Z}(M)$:
\begin{lemma} 
The Fourier transform $\mathcal{Z}(M)$ of the measure \eqref{measure4}
satisfies 
\begin{align}
\frac{1}{\mathrm{i}}\frac{\partial \mathcal{Z}(M)}{\partial M_{ab}} 
= \frac{\mathrm{i} M_{ba}\mathcal{Z}(M)}{N(E_a+E_b)} 
-\frac{\lambda }{\mathrm{i}^3(E_a+E_b)} 
\sum_{k,l=1}^{N} \frac{\partial^3
\mathcal{Z}(M)}{\partial M_{ak}\partial M_{kl}\partial M_{lb}}\;.
\label{eom}
\end{align}
\end{lemma}
\noindent\emph{Proof.} This follows from basic properties of 
the Gau\ss{}ian measure (\ref{measure0}). The 
derivative $\frac{1}{\mathrm{i}} 
\frac{\partial}{\partial M_{ab}}$ 
applied to $\mathcal{Z}(M)$ produces a factor $\Phi(e_{ab})$ under 
the integral. Moments of $d\mu_0(\Phi)$ are by 
(\ref{measure4}) a sum over pairings. This means that
$\Phi(e_{ab})$ is paired in all possible ways
with a $\Phi(e_{cd})$ contained in
$\exp(\mathrm{i} \Phi(M))$ or in $\mathcal{P}_4(\Phi,\lambda)$.
Every such pair contributes a factor 
$\frac{\delta_{ad}\delta_{bc}}{N(E_a+E_b)}$, and summing over
all pairings is the same as taking the derivative, thus producing a term 
\[
\frac{1}{N(E_a+E_b)}\Big(\mathrm{i} M_{ba}
-\lambda N
\sum_{k,l=1}^{N} \Phi(e_{ak})\Phi(e_{kl})\Phi(e_{lb})\Big)
\]
under the integral. The triple product of $\Phi(e_{..})$ is written as 
a third derivative with respect to the corresponding entries of $M$. 
\hspace*{\fill}$\square$%

\bigskip

The Kontsevich model
  \cite{Kontsevich:1992ti} with cubic deformation
  $\mathcal{P}_3(\Phi,\lambda)$ is governed by the
  equation of motion 
\[
\frac{1}{\mathrm{i}}\frac{\partial
    \mathcal{Z}(M)}{\partial M_{ab}} = \frac{\mathrm{i}
    M_{ba}\mathcal{Z}(M)}{N(E_a+E_b)} -\frac{\lambda
  }{\mathrm{i}^2 (E_a+E_b)} \sum_{k=1}^{N}
  \frac{\partial^2 \mathcal{Z}(M)}{\partial M_{ak} \partial M_{kb}}
  \:.
\]
For $N=1$ this is essentially the ODE
 $$f''(x)+2c f'(x)=x f(x)$$ solved by the Airy function $f(x)=e^{-cx}
  \mathrm{Ai}(x+c^2)$, hence the title of \cite{Kontsevich:1992ti}.
  Its quartic analogue is the matrix version of the ODE $$f'''(x)+3c
  f'(x)=x f(x)\:,$$ which does not seem to have a name. The Airy
  function is the case $p=2$ of a larger class 
\[
\mathrm{Ai}_p(x)= \frac{1}{2\pi }
\int_{-\infty}^{\infty } dt\;e^{\mathrm{i} (\frac{t^{p+1}}{p+1}+xt)}
\] 
of higher Airy functions. As remarked in \cite[\S
4.3]{Kontsevich:1992ti}, they also give rise to \emph{higher matrix Airy
functions}. In particular, there is also a `quartic analogue' $p=3$ in
this class, which was studied in \cite{Itzykson:1992ya, Kristjansen:1994hn}. 
This matrix model does not seem to be
related to our `quartic analogue' of the Kontsevich model.

Another equation of motion will be necessary for the subsequent
work in \cite{Branahl:2020yru}:
\begin{lemma}
The Fourier transform $\mathcal{Z}(M)$ of the measure \eqref{measure4}
satisfies 
\begin{align}
\frac{1}{N} \frac{\partial \mathcal{Z}(M)}{\partial E_{a}} 
= \Big(\sum_{k=1}^N  
\frac{\partial^2}{\partial M_{ak} \partial M_{ka}}
+ \frac{1}{N}\sum_{k=1}^N  G_{|ak|}+\frac{1}{N^2} 
G_{|a|a|}\Big)\mathcal{Z}(M)\;.
\label{eom-2}
\end{align}
\end{lemma}
\noindent\emph{Proof.}
Application of $\frac{1}{N} \frac{\partial}{\partial E_a} 
- \sum_{k=1}^N \frac{\partial^2}{\partial M_{ak} \partial M_{ka}}
- \frac{1}{N} \sum_{k=1}^N \frac{1}{E_a+E_k}$ to the left hand side of 
(\ref{measure0}) yields zero so that it gives 
\begin{align*}
\frac{1}{N}\frac{\partial}{\partial E_a} 
\big(d\mu_0(\Phi)\big)
= d\mu(\Phi)\sum_{k=1}^N \Big(\frac{1}{N(E_a+E_k)}-\Phi(e_{ka})\Phi(e_{ak})\Big)
\end{align*}
when applying it to the right hand side. Apply this identity to 
(\ref{measure4}) to get 
\begin{align*}
\frac{1}{N} \frac{\partial}{\partial E_a} 
\big(d\mu_\lambda(\Phi)\big) &=  d\mu_\lambda(\Phi)
\sum_{k=1}^N
  \Big(\frac{1}{N(E_a+E_k)}-\Phi(e_{ka})\Phi(e_{ak})\Big) 
\nonumber
\\*[-0.5ex]
& - d\mu_\lambda(\Phi) \!\! \int_{H_{N}'} d\mu_\lambda(\Phi) \sum_{k=1}^N
  \Big(\frac{1}{N(E_a+E_k)}-\Phi(e_{ka})\Phi(e_{ak})\Big)\;.
\end{align*}
Multiplying with $e^{\mathrm{i}\Phi(M)}$ and integrating over
$H_N'$ gives with (\ref{calZ4}) the assertion.
\hspace*{\fill} $\square$

\bigskip

The equations of motion (\ref{eom}) and (\ref{eom-2}) induce
identities between cumulants. Some of them are derived in
Appendix~\ref{appB}, for others see \cite{Grosse:2012uv}.  Taking also
the grading by $(g,b)$ into account, one can establish a partial order
in the homogeneous building blocks $G^{(g)}_{\dots}$. The least
element is the planar two-point function $G_{|ab|}^{(0)}$, which is
the dominant part (at large $N$) of the cumulant of length 2 and cycle
type $(0,1)$ (i.e.\ one cycle $ab$ of length $2$).  It satisfies a
closed non-linear equation for it alone, given in (\ref{2pt})
below. Any other homogeneous building block of (\ref{eq:Ggbn})
satisfies an affine equation with inhomogeneity that depends only on
functions of strictly larger topological \emph{Euler characteristic}
$\chi=2-2g-b$, which are known by induction.  Similar recursive
systems have been identified in many areas of mathematics. Their
common universal structure has been axiomatised under the name
\emph{topological recursion} \cite{Eynard:2007kz}, since the
recursion is by the topological Euler characteristic.  Starting from a
few initial data called the \emph{spectral curve}, topological
recursion constructs a hierarchy of differential forms and understands
them as spectral invariants of the curve. A prominent example
is the Kontsevich model \cite{Kontsevich:1992ti} whose topological
recursion is described e.g.\ in \cite[\S 6]{Eynard:2016yaa}.  Other classes
of examples are the one- and two-matrix models \cite{Chekhov:2006vd},
Mirzakhani's recursions \cite{Mirzakhani:2006fta} for the volume of
moduli spaces of Riemann surfaces, and recursions in Hurwitz theory
\cite{Bouchard:2007hi} and Gromov-Witten theory
\cite{Bouchard:2007ys}.

\section{The planar two-point function}

The two-point function $G_{|ab|}$ is the cumulant of length 2 and 
cycle type $(0,1)$ (i.e.\ one cycle $ab$ of length $2$), see
Appendix~\ref{appA}. We reprove in Appendix \ref{appB} that the planar
two-point function $G^{(0)}_{|ab|}$ (of degree or genus $g=0$)
satisfies
\begin{align}
\Big(E_a+E_b+\frac{\lambda}{N} 
\sum_{k=1}^{N} G^{(0)}_{|ak|}\Big) 
G^{(0)}_{|ab|}
= 1+ \frac{\lambda}{N} \sum_{k=1}^{N} 
\frac{G^{(0)}_{|kb|}-G^{(0)}_{|ab|}}{E_k-E_a}\;.
\label{2pt}
\end{align}
This equation was first established in \cite{Grosse:2009pa};   
equation (\ref{DS2-final}) which involves all $G^{(g)}_{|ab|}$ was
obtained in \cite{Grosse:2012uv}.

To give a meaning to the term $k=a$ in (\ref{2pt}) we make the
decisive assumption that $\{G^{(0)}_{|ab|}\}_{a,b=1,\dots N}$
arise by evaluation of a holomorphic function in two complex
variables. Let $E_1,\dots,E_d$ be the distinct 
entries in the tuple $(E_k)$, which occur with multiplicities 
$r_1,\dots,r_d$, with $N=r_1+\dots+r_d$. We assume that for some neighbourhoods 
$\mathcal{U}_k\subset \mathbb{C}$ of $E_k$ there is a 
holomorphic function 
$G^{(0)}: \bigcup_{k,l=1}^d (\mathcal{U}_k\times \mathcal{U}_l) 
\to \mathbb{C}$ which interpolates 
$G^{(0)}_{|ab|}=G^{(0)}(E_a,E_b)$ and satisfies the natural (but by
no means unique) holomorphic extension 
\begin{align}
\Big(\zeta+\eta+\frac {\lambda}{N}
\sum_{k=1}^d r_k G^{(0)}(\zeta,E_k)\Big)G^{(0)}(\zeta,\eta) 
&=1+ 
\frac{\lambda}{N} \sum_{k=1}^d r_k 
\frac{G^{(0)}(E_k,\eta)-G^{(0)}(\zeta,\eta)}{E_k-\zeta}
\label{eq:GZW}
\end{align}
of (\ref{2pt}), for 
$(\zeta,\eta)\in \bigcup_{k,l=1}^d (\mathcal{U}_k\times \mathcal{U}_l)$. 
Equation (\ref{2pt}) is understood as the
limit $\zeta\to E_a$ and $\eta\to E_b$ of (\ref{eq:GZW}) when 
taking multiplicities into account. It is not possible
to deduce (\ref{eq:GZW}) from (\ref{2pt}) alone.
Justification of (\ref{eq:GZW}) comes from the fact
that it gives rise to interesting mathematical structures:
\begin{theorem}
\label{thm:main}
Construct $2d$ functions $\{\varepsilon_k(\lambda),
\varrho_k(\lambda)\}_{k=1,\dots,d}$,
with $\lim_{\lambda\to 0} (\varepsilon_k,\varrho_k)=(E_k,r_k)$,
as implicitly defined solution of 
the system
\begin{align}
E_k=R(\varepsilon_k)\;,\qquad
r_k=\varrho_k R'(\varepsilon_k)\;,\qquad
\text{where}\quad
R(z)=z-\frac{\lambda}{N} 
\sum_{j=1}^d \frac{\varrho_j}{\varepsilon_j+z}\;.
\label{thm:R}
\end{align}\enlargethispage{0.9mm}%
Then \eqref{eq:GZW} is solved by $G^{(0)}(\zeta,\eta)
=\mathcal{G}^{(0)}(R^{-1}(\zeta),R^{-1}(\eta))$,
where $\mathcal{G}^{(0)}:\hat{\mathbb{C}}\times \hat{\mathbb{C}} \to 
\hat{\mathbb{C}}$ is the rational function
\begin{align}
\mathcal{G}^{(0)}(z,w)
&=\frac{1}{(R(w)-R(-z))(R(z)-R(-w))}
\Big\{R(z)+R(w)
\nonumber
\\*
&+\frac{\lambda}{N}
\sum_{k=1}^d \Big(\frac{r_k}{R(\varepsilon_k)-R(z)}
+\frac{r_k}{R(\varepsilon_k)-R(w)}\Big)
\nonumber
\\*
&+\frac{\lambda^2}{N^2}
\sum_{k,l=1}^d \frac{r_k r_l \mathcal{G}_{kl}}{
(R(\varepsilon_k)-R(z))(R(\varepsilon_l)-R(w))}
\Big\}
\label{Gzw-new}
\end{align}
with 
\begin{align}
\mathcal{G}_{kl}
=\frac{\displaystyle
\Bigg(\prod_{j,m=1}^d \frac{(-\widehat{\varepsilon_k}^j
-\widehat{\varepsilon_l}^m)}{\varepsilon_j+\varepsilon_m}\Bigg)
\Bigg(
\prod_{\substack{j=1\\j\neq k}}^d \frac{\varepsilon_k-\varepsilon_j}
{R(\varepsilon_k){-}R(\varepsilon_j)}\Bigg)
\Bigg(
\prod_{\substack{m=1\\m\neq l}}^d \frac{\varepsilon_l-\varepsilon_m}
{R(\varepsilon_l){-}R(\varepsilon_m)}\Bigg)
}{R'(\varepsilon_k)R'(\varepsilon_l) (\varepsilon_k+\varepsilon_l)} \;.
\label{Ekl}
\end{align}
Here $z\in \{u,\hat{u}^1,\dots,\hat{u}^d\}$ is the list of the
different roots of $R(z)=R(u)$, and the correct branch of $R^{-1}$ is
chosen by the implicitly defined solutions above (i.e.\ 
$\varepsilon_k \in R^{-1}(\mathcal{U}_k)$ for this branch).
In particular, $G^{(0)}_{|kl|}
=\mathcal{G}^{(0)}(\varepsilon_k,\varepsilon_l)\equiv 
\mathcal{G}_{kl}$ solves \eqref{2pt}. 
\end{theorem}

Existence of $(\varepsilon_k(\lambda),\varrho_k(\lambda))$ in
a neighbouhood of $\lambda=0$ is
guaranteed by the implicit function theorem. We will prove several
equivalent formulae for $\mathcal{G}^{(0)}(z,w)$: (\ref{calG:zhatw}),
(\ref{Gzw-symm}), (\ref{Gzw-final}) and eventually
(\ref{Gzw-new}). Some of them were already proved in
\cite{Grosse:2019jnv}.  There, inspired by the solution of a
particular case \cite{Panzer:2018tvy}, equation (\ref{eq:GZW}) was
interpreted as an integral equation for a Dirac measure. Approximating
the Dirac measure by a H\"older-continuous function allowed to employ
boundary values techniques for sectionally holomorphic functions.
Residue theorem and Lagrange-B\"urmann resummation gave a solution
formula whose limit back to Dirac measure was arranged into
(\ref{Gzw-final}).

In this paper we provide a more elementary proof of these equations 
which solely needs properties of
\emph{Cauchy matrices} established by Schechter
\cite{Schechter:1959??}:
\begin{proposition}[\cite{Schechter:1959??}]
\label{prop:Schechter}
For two $d$-tuples $(a_1,\dots,a_d)$ and $(b_1,\dots,b_d)$, with 
all $a_i,b_j$ distinct,
consider the $d\times d$-matrix $H=(\frac{1}{a_k-b_l})_{kl}$. 
Let $A(x):=\prod_{i=1}^d (x-a_i)$ and 
$B(y):=\prod_{j=1}^d (y-b_j)$. Then the
inverse of $H$ is given by
\begin{align}
(H^{-1})_{kl}= (a_l-b_k) A_l(b_k)\,B_k(a_l)\;,
\end{align}
where $A_l,B_k$ are the Lagrange interpolation polynomials 
\begin{align}
A_l(x)=\frac{A(x)}{(x-a_l) \,A'(a_l)}\;,\qquad
B_k(y)=\frac{B(y)}{(y-b_k) \,B'(b_k)}\;.
\end{align}
The inverse of $H$ satisfies 
\begin{align}
\frac{B_k(x)A(b_k)}{A(x)}=\sum_{l=1}^d \frac{(H^{-1})_{kl}}{a_l-x}\;,\qquad
\frac{A_l(x)B(a_l)}{B(x)}=\sum_{k=1}^d \frac{(H^{-1})_{kl}}{x-b_k}\;.
\label{Schechter12}
\end{align}
Moreover, the row sums and column sums of $H^{-1}$ are given by 
\begin{align}
\sum_{j=1}^d (H^{-1})_{kj}=
-\frac{A(b_k)}{B'(b_k)}\;,\qquad
\sum_{i=1}^d (H^{-1})_{il}=
\frac{B(a_l)}{A'(a_l)}\;,
\label{Schechter:row}
\end{align}
and one has, for all $j=1,\dots, d$, 
\begin{align}
\sum_{k=1}^d \frac{A(b_k)}{(b_k-a_j) B'(b_k)}=1
\quad\text{and}\quad
\sum_{l=1}^d \frac{B(a_l)}{(a_l-b_j) A'(a_l)}
=1\;.
\label{Schechter-res}
\end{align}
\end{proposition}

\section{Proof of Theorem~\ref{thm:main}}

We are going to construct a non-constant rational function $R\in
\mathbb{C}(z)$ viewed as a branched cover
$R:\hat{\mathbb{C}}=\mathbb{C}\cup\{\infty\}\to
\hat{\mathbb{C}}=\mathbb{C}\cup\{\infty\}$ of Riemann surfaces (with
$z=\mathrm{id}_{\mathbb{C}}$ the standard coordinate on $\mathbb{C}$) via the
following:
\begin{ansatz}
\label{ansatz}
A branched cover $R:\hat{\mathbb{C}}\to 
\hat{\mathbb{C}}$ is supposed to be determined by 
conventions \textup{(i)--(vi)} and an algebraic relation
\textup{(vii)}:
\begin{enumerate}[\rm(i)]
\item $R$ has degree $d+1$.
\item All ramification points of $R$ do not belong to  $R^{-1}(\{E_1,\dots,E_d\})$.
\item Without loss of generality, $R(\infty)=\infty$ with residue $-1$
  in the sense that $ \Res_{z= \infty} R(z)dz=-1$.

\item For every $k=1,\dots, d$, distinguish any of the $d{+}1$ distinct
  points of the fibre $R^{-1}(E_k)$ as $\varepsilon_k$.  Take any
  connected neighbourhood $\mathcal{U}_k\subset \mathbb{C}$ of $E_k$
  for which $R^{-1}(\mathcal{U}_k)$ has $d{+}1$ connected components,
  and let $\mathcal{V}_k$ be the connected component of
  $R^{-1}(\mathcal{U}_k)$ which contains $\varepsilon_k$.  Then the
  choice of $\{\varepsilon_k\}$ determines a holomorphic function
  $\mathcal{G}^{(0)}:\bigcup_{k,l=1}^d (\mathcal{V}_k\times
  \mathcal{V}_l) \to \mathbb{C}$ by
  $\mathcal{G}^{(0)}(z,w)=G^{(0)}(R(z),R(w))$, where $G^{(0)}: 
\bigcup_{k,l=1}^d (\mathcal{U}_k\times \mathcal{U}_l) 
\to \mathbb{C}$   satisfies \eqref{eq:GZW}.

\item For any $w\in \bigcup_{j=1}^d\mathcal{V}_j$, let
  $\hat{w}^1,\dots,\hat{w}^d$ be the $d$ other distinct preimages of
  $R(w)\in \bigcup_{j=1}^d\mathcal{U}_j$ under $R$, i.e.\
  $R(\hat{w}^k)=R(w)$. Assume that
\[
\infty\neq R(-\hat{w}^l) \quad \text{and} \quad  R(-\hat{w}^l) \neq
R(-\hat{w}^{l'})
\]
for all $l, l'=1,\dots,d$ with $l\neq l'$ and $w$ close to some
$\varepsilon_k$.

\item For any $w$ close to some $\varepsilon_k$, 
$\mathcal{G}^{(0)}(-\hat{w}^l,w)$ is defined and finite 
for all $l=1,\dots,d$. This is the case   
e.g.\ if $-\hat{w}^l\in \bigcup_{j=1}^d\mathcal{V}_j$ for all 
$l=1,\dots,d$, or if $\mathcal{G}^{(0)}$ extends to a suitable 
rational function on
$\hat{\mathbb{C}}\times \hat{\mathbb{C}}$.

\item For any $z\in \mathcal{V}_l$ one has
\begin{align}
\fbox{$\displaystyle R(z)+\frac{\lambda}{N}\sum_{k=1}^d
r_k \mathcal{G}^{(0)}(z,\varepsilon_k)
+ \frac{\lambda}{N} \sum_{k=1}^d \frac{r_k}{R(\varepsilon_k)-R(z)}
=-R(-z)$}
\label{Rznegz}
\end{align}
\end{enumerate}
\end{ansatz}

With the properties (iv) and (vii) in this Ansatz~\ref{ansatz} 
we turn (\ref{eq:GZW}) into 
\begin{align}
\big(R(w)-R(-z)\big)\mathcal{G}^{(0)}(z,w) =1+ 
\frac{\lambda}{N} \sum_{k=1}^d r_k\frac{
\mathcal{G}^{(0)}(\varepsilon_k,w)}{
R(\varepsilon_k)-R(z)}\;,
\label{GRR}
\end{align}
where $(z,w)\in \bigcup_{k,l=1}^d (\mathcal{V}_k\times
\mathcal{V}_l)$.  Next, setting $z=-\hat{w}^l$ in (\ref{GRR}) for
$l=1,\dots,d$ and a given $w$ close to some $\varepsilon_k$,
requirements (v) and (vi) of Ansatz~\ref{ansatz} give (by $\infty\neq
R(-\hat{w}^l)$ and $\mathcal{G}^{(0)}(-\hat{w}^l,w)$ is defined and
finite) the $d$ equations
\begin{align}
\frac{\lambda}{N} 
\sum_{k=1}^d r_k \frac{\mathcal{G}^{(0)}(\varepsilon_k,w)}{
R(-\hat{w}^l)-R(\varepsilon_k)}=1\;. 
\end{align}
This identifies $\frac{\lambda}{N} r_k
\mathcal{G}^{(0)}(\varepsilon_k,w)$ as row sums of the inverse of a
Cauchy matrix.  Setting $a_j=R(-\hat{w}^j)$ and $b_i=R(\varepsilon_i)$
in the first identity (\ref{Schechter:row}) in 
Proposition~\ref{prop:Schechter} we conclude (since the $a_j,b_i$
for $j,i=1,\dots,d$ are pairwise distinct by requirement (v) of
Ansatz~\ref{ansatz}):
\begin{corollary}
With \textup{Ansatz \ref{ansatz}} 
one has
\begin{align}
\frac{\lambda}{N} r_k \mathcal{G}^{(0)}(\varepsilon_k,w)
= -\frac{\prod_{j=1}^d (R(\varepsilon_k)-R(-\hat{w}^j))}
{\prod_{j=1,j\neq k}^d (R(\varepsilon_k)-R(\varepsilon_j))}\;.
\label{calG:epshatw}
\end{align}
\end{corollary}
\noindent
Inserted back into (\ref{GRR}) expresses $\mathcal{G}^{(0)}(z,w)$ in 
terms of $R$. The result simplifies:
\begin{lemma}
\label{lemma:Gzw}
With \textup{Ansatz \ref{ansatz}} one has
\begin{align}
\mathcal{G}^{(0)}(z,w)=\frac{1}{(R(w)-R(-z))} \prod_{j=1}^d
\frac{R(z)-R(-\hat{w}^j)}{R(z)-R(\varepsilon_j)}\;.
\label{calG:zhatw}
\end{align}
\end{lemma}
\noindent\emph{Proof.} This follows from (\ref{Schechter-res}) for 
$(d+1)$-tuples with index $0$ prepended.  
Setting $b_0=R(z)$, $b_k=R(\varepsilon_k)$, $a_l=R(-\hat{w}^l)$ for 
$k,l=1,\dots,d$, then the case $j=0$ of the first identity 
(\ref{Schechter-res}) reads (independent of $a_0$)
\begin{align}
\frac{\prod_{j=1}^d (R(z)-R(-\hat{w}^j))}{\prod_{j=1}^d
  (R(z)-R(\varepsilon_j))}
+ \sum_{k=1}^d \frac{\prod_{j=1}^d (R(\varepsilon_k)
-R(-\hat{w}^j))}{(R(\varepsilon_k)-R(z)) 
\prod_{j=1,j\neq k}^d  (R(\varepsilon_k)-R(\varepsilon_j))}=1\;.
\label{identity}
\end{align}
Equation (\ref{calG:zhatw}) results from this identity when 
inserting (\ref{calG:epshatw}) into (\ref{GRR}).
\hspace*{\fill}$\square$%

\begin{lemma}
\label{lemma:R}
With \textup{Ansatz \ref{ansatz}} the rational function  $R\in\mathbb{C}(z)$
is necessarily given by 
\begin{align}
R(z)=z+c_0-\frac{\lambda}{N} 
\sum_{k=1}^d \frac{\varrho_k}{\varepsilon_k+z}
\quad\text{for some } c_0\in \mathbb{C}\;,\qquad 
\varrho_k=\frac{r_k}{R'(\varepsilon_k)}\;.
\label{eq:R}
\end{align}
\end{lemma}
\noindent\emph{Proof.}
Comparing the limit $z\to \varepsilon_k$ of (\ref{calG:zhatw}) with 
(\ref{calG:epshatw}) shows that $R$ has a simple pole at every
$-\varepsilon_k$ with 
\[
R'(\varepsilon_k) \Res\displaylimits_{z= -\varepsilon_k} R(z) dz=
-\frac{\lambda r_k}{N} \neq 0 \:.
\] 
By construction, $R$ has also a pole at $\infty$. Since
$R$ has degree $d+1$ by (i) in \textup{Ansatz \ref{ansatz}},
$\{-\varepsilon_1,\dots, -\varepsilon_d,\infty\}$ is already the
complete list of poles (i.e.\ preimages of $\infty$) of $R$. Moreover,
the pole at $\infty$ has to be simple with
$\lim_{z\to\infty}\frac{R(z)}{z}=1$ by (iii) in \textup{Ansatz
  \ref{ansatz}}.  Therefore, $R(z)-z+\frac{\lambda}{N} \sum_{k=1}^d
\frac{\varrho_k}{\varepsilon_k+z}$ is a bounded holomorphic function
on $\hat{\mathbb{C}}$, which by Liouville's theorem is a constant
$c_0$.\hspace*{\fill}$\square$%

\begin{corollary}
  For $u\in \bigcup_{j=1}^d\mathcal{V}_j$ one has an equality of
  rational functions in $z$:
\begin{align}
R(z)-R(u)=(z-u) \prod_{k=1}^d \frac{z-\hat{u}^k}{z+\varepsilon_k}\;,
\label{R-rational}
\end{align}
where $\hat{u}^k$ are the other preimages 
of $R(u)$ under $R$.
\end{corollary}
\noindent\emph{Proof.}
Both sides are a rational function $r(z)$, with zeros only in
$u,\hat{u}^1,\dots,\hat{u}^d$ and poles only in
$-\varepsilon_1,\dots,-\varepsilon_d,\infty$, all of which are simple. So
they differ by a constant factor, which has to be $1$ because both
sides satisfy $\lim_{z\to\infty}\frac{r(z)}{z}=1$.  \hspace*{\fill}$\square$%
\begin{proposition}
\label{prp:symm}
With \textup{Ansatz \ref{ansatz}} the two-point function is symmetric, 
$\mathcal{G}^{(0)}(z,w)=
\mathcal{G}^{(0)}(w,z)$. One has $\mathcal{G}^{(0)}(\varepsilon_k,
\varepsilon_l)=\mathcal{G}_{kl}$ with 
$\mathcal{G}_{kl}$ given in \eqref{Ekl}.
\end{proposition}
\noindent\emph{Proof.}
Inserting (\ref{R-rational}) into (\ref{calG:zhatw}) gives for 
$z,w\in \bigcup_{j=1}^d\mathcal{V}_j$:
\begin{align}
\mathcal{G}^{(0)}(z,w)&=\frac{\prod_{k=1}^d(\varepsilon_k-z)}{
(z+w) \prod_{k=1}^d (-z-\hat{w}^k)}
\prod_{k=1}^d \frac{\displaystyle
\frac{(z+\hat{w}^k)\prod_{l=1}^d (-\hat{w}^k-\hat{z}^l)}{
\prod_{l=1}^d (\varepsilon_l-\hat{w}^k)} }{
\displaystyle \frac{
(z-\varepsilon_k)\prod_{l=1}^d (\varepsilon_k-\hat{z}^l)}{
\prod_{l=1}^d(\varepsilon_k+\varepsilon_l)}
}
\nonumber
\\[-1ex]
&= \frac{1}{(z+w)} \prod_{k,l=1}^d 
\frac{(\varepsilon_k+\varepsilon_l)(-\hat{w}^k-\hat{z}^l)}{
(\varepsilon_k-\hat{z}^l)(\varepsilon_l-\hat{w}^k)}
\nonumber
\\
&= \frac{1}{(z+w)} \prod_{k,l=1}^d 
\frac{(-\hat{w}^k-\hat{z}^l)}{(\varepsilon_k+\varepsilon_l)}
\frac{(\varepsilon_k-z)}{(R(\varepsilon_k)-R(z))}
\frac{(\varepsilon_l-w)}{(R(\varepsilon_l)-R(w))}
\;.
\label{Gzw-symm}
\end{align}
The limit $z\to \varepsilon_k$ and $w\to \varepsilon_l$ gives 
$\mathcal{G}^{(0)}(\varepsilon_k,
\varepsilon_l)=\mathcal{G}_{kl}$. \hspace*{\fill} $\square$

\bigskip

We prove that (ii),\,(iv),\,(v),\,(vi) and (vii) of Ansatz \ref{ansatz} are 
automatic. We start with
\begin{proposition}
\label{prop:vii}
Relation \textup{(vii)} of \textup{Ansatz \ref{ansatz}} 
is consistent provided that $c_0=0$.
\end{proposition}
\noindent\emph{Proof.} 
With Lemma~\ref{lemma:R} and Lemma~\ref{lemma:Gzw}, both a consequence
of \textup{Ansatz \ref{ansatz}}, each side of (\ref{Rznegz}) is a
rational function, and all poles are simple. For the term
$\frac{r_k}{R(\varepsilon_k)-R(z)}$ this follows from the assumption
(ii) of \textup{Ansatz \ref{ansatz}}.  We show that both sides of
(\ref{Rznegz}) have the same simple poles with the same residues. Then
by Liouville's theorem their difference is a constant, which is easy
to control.
 
First, it follows from (\ref{eq:R}) and (\ref{calG:zhatw}) that both
sides of (\ref{Rznegz}) approach $z$ for $z\to \infty$. Near $\infty$
the difference between both sides of (\ref{Rznegz}) is $\pm 2c_0$,
which shows that $c_0=0$ in (\ref{eq:R}) is necessary.

Next, (\ref{eq:R}) shows that the only other poles of the right hand
side of (\ref{Rznegz}) are simple and located at $z=\varepsilon_k$
with residue $-\frac{\lambda}{N} \varrho_k$. The same simple poles
with the same residues are produced by $\frac{\lambda}{N} \sum_{k=1}^d
\frac{r_k}{R(\varepsilon_k)-R(z)}$ on the left hand side, taking
$r_k/R'(\varepsilon_k)=\varrho_k$ into account.

But the left hand side of (\ref{Rznegz}) could also have poles 
at $z=-\varepsilon_j$
and $z=\widehat{\varepsilon_m}^j$ 
(see (\ref{eq:R}) and (\ref{Gzw-symm})).  We have $\Res_{z=
  -\varepsilon_j} R(z)dz= -\frac{\lambda}{N} \varrho_j$. Setting
$w\mapsto \varepsilon_l$ in (\ref{calG:zhatw}), then with $\lim_{z\to
  -\varepsilon_j} \frac{R(z)-R(-\widehat{\varepsilon_l}^k)}{
  R(z)-R(\varepsilon_k)}=1$ for any $k,l$ (here 
\textup{(v)} of \textup{Ansatz \ref{ansatz}} is used)
one easily finds that
$\mathcal{G}^{(0)}(-\varepsilon_j,\varepsilon_l)$ is finite for
$j\neq l$ and that $\Res_{z= -\varepsilon_j} \frac{\lambda}{N}
r_j\mathcal{G}^{(0)}(z,\varepsilon_j)dz= \frac{\lambda}{N}
\frac{r_j}{R'(\varepsilon_j)}$, which thus cancels $\Res_{z=
  -\varepsilon_j} R(z)dz= -\frac{\lambda}{N} \varrho_j$.

Finally, from (\ref{calG:zhatw}) we conclude 
\begin{align*}
\Res\displaylimits_{z= \widehat{\varepsilon_m}^j}
\mathcal{G}^{(0)}(z,\varepsilon_k)dz 
&=\frac{1}{(R(\varepsilon_k)
-R(-\widehat{\varepsilon_m}^j)) R'(\widehat{\varepsilon_m}^j)} 
\frac{\prod_{i=1}^d 
(R(\varepsilon_m)-R(-\widehat{\varepsilon_k}^i))}{
\prod_{i=1,i\neq m }^d (R(\varepsilon_m)-R(\varepsilon_i))}
\\
&=\frac{1}{(R(\varepsilon_k)
-R(-\widehat{\varepsilon_m}^j)) R'(\widehat{\varepsilon_m}^j)} 
\Big( - \frac{\lambda}{N} r_m 
\mathcal{G}^{(0)}(\varepsilon_m,\varepsilon_k)\Big)
\\
&= \frac{r_m}{r_k R'(\widehat{\varepsilon_m}^j)} 
\frac{1}{(R(\varepsilon_k)
-R(-\widehat{\varepsilon_m}^j)) } 
\Big( - \frac{\lambda}{N} r_k 
\mathcal{G}^{(0)}(\varepsilon_k,\varepsilon_m)\Big)
\\
&= \frac{r_m}{r_k R'(\widehat{\varepsilon_m}^j)} 
\frac{1}{(R(\varepsilon_k)
-R(-\widehat{\varepsilon_m}^j))} 
\frac{\prod_{i=1}^d 
(R(\varepsilon_k)-R(-\widehat{\varepsilon_m}^i))}{
\prod_{i=1,i\neq k }^d (R(\varepsilon_k)-R(\varepsilon_i))}\;,
\end{align*}
where (\ref{calG:epshatw}), the symmetry 
$\mathcal{G}^{(0)}(\varepsilon_m,\varepsilon_k)=
\mathcal{G}^{(0)}(\varepsilon_k,\varepsilon_m)$ and 
again (\ref{calG:epshatw}) have been used. The first identity 
(\ref{Schechter-res}) for $b_k=R(\varepsilon_k)$ and 
$a_j=R(-\widehat{\varepsilon_m}^j)$ gives 
\[
\Res\displaylimits_{z= \widehat{\varepsilon_m}^j}
\sum_{k=1}^d r_k \mathcal{G}^{(0)}(z,\varepsilon_k)dz=
\frac{r_m}{R'(\widehat{\varepsilon_m}^j)}\;,
\] 
which precisely cancels 
$\Res_{z= \widehat{\varepsilon_m}^j}
\sum_{k=1}^d
\frac{r_k}{R(\varepsilon_k)-R(z)}dz
= -\frac{r_m}{R'(\widehat{\varepsilon_m}^j)}
$. \hspace*{\fill}$\square$

\bigskip

Let us consider now the rational function 
\begin{equation}\label{eq2:R}
R(z)=z-\frac{\lambda}{N} 
\sum_{k=1}^d \frac{\varrho_k}{\varepsilon_k+z}
\end{equation}
from equation (\ref{eq:R}) with $c_0=0$. We are interested in the real
and complex solutions $\{\varepsilon_k,\varrho_k\}_{k=1,\dots,d}$
(depending on $\lambda$) of the $2d$ equations
\begin{subequations}
\begin{align}
\label{eq:f-l}
0&=R(\varepsilon_l)-E_l=\varepsilon_l -E_l -\frac{\lambda}{N} 
\sum_{k=1}^d \frac{\varrho_k}{\varepsilon_k+\varepsilon_l}  
=:f_l(\varepsilon_1,\varrho_1,\cdots,\varepsilon_d,\varrho_d,\lambda)\;,
\\
\label{eq:g-l}
0&= R'(\varepsilon_l)- \frac{r_l}{\varrho_l}=1-\frac{r_l}{\varrho_l}+ \frac{\lambda}{N} 
\sum_{k=1}^d \frac{\varrho_k}{(\varepsilon_k+\varepsilon_l)^2}=:g_l(\varepsilon_1,\varrho_1,\cdots,\varepsilon_d,\varrho_d,\lambda)
\end{align}
\end{subequations}
for $l=1,\dots,d$ from Theorem \ref{thm:main} for the given positive
real numbers $E_l>0$ and $r_l>0$ from Section
\ref{sec:matrixintegrals}. In the following we only use that all these
$E_l,r_l$ are positive (but not that $r_l$ is an integer counting the
multiplicity of $E_l$).  So we consider for $\mathbb{K}=\mathbb{R}$ or
$\mathbb{K}=\mathbb{C}$ the real or complex algebraic subset
$$ Z(\mathbb{K}):=\{(\varepsilon_1,\varrho_1,\cdots,\varepsilon_d,\varrho_d,\lambda)\in U(\mathbb{K})|\: f_l=0,
g_l=0,\: l=1,\dots,d\}$$
in 
\[
U(\mathbb{K}):=\{(\varepsilon_1,\varrho_1,\cdots,\varepsilon_d,
\varrho_d,\lambda)|\:
\varepsilon_k+\varepsilon_l\neq 0, \: \varrho_l\neq 0,\:
k,l=1,\dots,d\}\subset \mathbb{K}^{2d+1}\:,
\]
i.e.\ in the complement of the corresponding central hyperplane
arrangement in $\mathbb{K}^{2d+1}$.  Note that
$(\varepsilon_1,\varrho_1,\cdots,\varepsilon_d,\varrho_d,0)\in
Z(\mathbb{K})$ iff $\varepsilon_l=E_l$ and $\varrho_l=r_l$ for all $
l=1,\dots,d$, and this real point belongs to the chamber
\[
U_{+}(\mathbb{R}):=\{(\varepsilon_1,\varrho_1,\cdots,\varepsilon_d,
\varrho_d,\lambda)|\:
\varepsilon_l > 0, \: \varrho_l > 0,\:  l=1,\dots,d\}\subset
U(\mathbb{R})\subset \mathbb{R}^{2d+1}\:.
\]

Note that the complex dimension of any irreducible component of
$Z(\mathbb{C})$ is at least $1=(2d+1)-2d$, since we are considering
$2d$ equations in a Zariski-open subset of $ \mathbb{C}^{2d+1}$.

\begin{lemma}
  $Z(\mathbb{K})$ is a one-dimensional (real or complex) algebraic
  submanifold of $U(\mathbb{K})$ near the reference point
  $(E_1,r_1,\cdots, E_d,r_d,0)$, with the projection
\[
pr: \mathbb{K}^{2d+1} \supset  Z(\mathbb{K}) \to \mathbb{K}
\]
onto the last $\lambda$-coordinate a submersion near $(E_1,r_1,\cdots,
E_d,r_d,0)$ in the Zariski topology. In particular, the reference
point $(E_1,r_1,\cdots, E_d,r_d,0)$ only belongs to one irreducible
component of $Z(\mathbb{C})$, which is of dimension one.
Moreover, the map (of pointed sets)
$$pr: Z(\mathbb{K}) \to \mathbb{K} \quad \text{ with}  \quad pr(E_1,r_1,\cdots, E_d,r_d,0)=0$$
becomes locally near $(E_1,r_1,\cdots, E_d,r_d,0)$ a (real or complex)
analytic isomorphism onto an open interval or disc around
$\lambda=0\in \mathbb{K}$, fitting with the description given in
Theorem~\textup{\ref{thm:main}} in terms of the implicit function theorem.
\end{lemma}

\noindent\emph{Proof.}
The claim follows from
\begin{align*}
\Big(\frac{\partial f_l}{\partial \varepsilon_k}(E_1,r_1,\cdots, E_d,r_d,0), \frac{\partial f_l}{\partial \varrho_k}(E_1,r_1,\cdots, E_d,r_d,0)\Big)
&=(\delta_{lk},0)\\
\text{and}\quad
\Big(\frac{\partial g_l}{\partial \varepsilon_k}(E_1,r_1,\cdots, E_d,r_d,0), \frac{\partial g_l}{\partial \varrho_k}(E_1,r_1,\cdots, E_d,r_d,0)\Big)
&=\Big(0, \delta_{lk}\cdot \frac{1}{r_l}\Big)
\end{align*}
for $l,k=1,\dots,d$.
\hfill $\square$%

\begin{remark}
  Let us rewrite for a fixed $\lambda\in \mathbb{C}$ the equations
  (\ref{eq:f-l}) and (\ref{eq:g-l}) in terms of the $d$ polynomials
\[
F_l(\varepsilon_1,\varrho_1,\cdots,\varepsilon_d,\varrho_d) :=
f_l(\varepsilon_1,\varrho_1,\cdots,\varepsilon_d,\varrho_d,\lambda)
\cdot \prod_{k=1}^d(\varepsilon_k+\varepsilon_l)
\]
of degree $d+1$ and  the $d$  polynomials
\[
G_l(\varepsilon_1,\varrho_1,\cdots,\varepsilon_d,\varrho_d) :=
g_l(\varepsilon_1,\varrho_1,\cdots,\varepsilon_d,\varrho_d,\lambda)\cdot
\varrho_l \prod_{k=1}^d(\varepsilon_k+\varepsilon_l)^2
\]
of degree $2d+1$ (for $l=1,\dots,d$), so that:
\[
Z(\mathbb{C})\cap\{pr=\lambda\} \subset 
\{(\varepsilon_1,\varrho_1,\cdots,\varepsilon_d,\varrho_d)
\in \mathbb{C}^{2d}|\: F_l=0,
G_l=0,\: l=1,\dots,d\}\times\{\lambda\} \:.
\]
If for a given $\lambda\in \mathbb{C}$ the set 
\[
\{(\varepsilon_1,\varrho_1,\cdots,\varepsilon_d,\varrho_d)
\in \mathbb{C}^{2d}|\: F_l=0,
G_l=0,\: l=1,\dots,d\}
\] 
is finite, then one gets by the \emph{affine
  Bezout inequality} \cite[Thm 3.1]{affineBezout} the upper estimate
$(d+1)^d(2d+1)^d$ for the number of solutions of the equations
(\ref{eq:f-l}) and (\ref{eq:g-l}) (for this $\lambda$).
\end{remark}

Let us come back to the rational function $R(z)$ from (\ref{eq2:R}) for the case of positive real $E_l>0$ and $r_l>0$ 
for $ l=1,\dots,d$ related to the solution of  Theorem~\textup{\ref{thm:main}} as discussed before. Then
\begin{equation}
R'(z)=1+\frac{\lambda}{N} 
\sum_{k=1}^d \frac{\varrho_k}{(\varepsilon_k+z)^2}>0
\end{equation}
for all $\lambda\geq 0$ and $z\in \mathbb{R}\backslash \{-\varepsilon_1,\dots,-\varepsilon_d\}$.

\begin{lemma}
\label{lemma:neighbourhood}
  $R^{-1}(E_k)$ consists for all $E_k>0$ of $d+1$ different real
  points so that assumption \textup{(ii)} of \textup{Ansatz \ref{ansatz}} holds.  Moreover we can choose
  $\mathcal{U}=\mathcal{U}_1=\cdots=\mathcal{U}_d$ as a small simply
  connected open neighbourhood of $(R(0),\infty)\subset \mathbb{R}$ in $\mathbb{C}$, with
  $\mathcal{V}=\mathcal{V}_1= \cdots= \mathcal{V}_d$ also containing
  $(0,\infty)$. By shrinking of $\mathcal{U}$ we can even assume that $\mathcal{V}\subset \{z\in \mathbb{C}|\;\mathrm{Re}(z)>0\}$.
  Then the assumptions \textup{(v)} and 
\textup{(vi)} of
  \textup{Ansatz \ref{ansatz}} hold for all $w\in \mathcal{V}$ with
  $\mathcal{V}$ small enough, as well as the assumption \textup{(iv)} with $\mathcal{G}^{(0)}$ as in \textup{(\ref{calG:zhatw})}
  resp.\ \textup{(\ref{Gzw-symm})}.
\end{lemma}

\noindent\emph{Proof.}
 If we order the numbering of the $\varepsilon_l$ as
$\varepsilon_i<\varepsilon_{i+1}$ for $i=0,\dots,d$ with $\varepsilon_0:=-\infty$ and $\varepsilon_{d+1}:=+\infty$, then
$$R: (-\varepsilon_{i+1},-\varepsilon_i)\to \mathbb{R}$$ 
is for all $i=0,\dots,d$ strictly monotone increasing and bijective by
the estimate $R'(z)>0$ above and the intermediate value theorem. This
proves the first claim. Similarly, $R^{-1}(R(w))$ consists for any 
$w\in\mathcal{V}\cap (0,\infty)$ of $d+1$ real points which we can order as  
$\hat{w}^l\in (-\varepsilon_{l+1},-\varepsilon_l)$. Therefore, 
$-\hat{w}^l\in (0,\infty)$ and all $R(-\hat{w}^l)$ are distinct for
$w\in\mathcal{V}\cap (0,\infty)$, because $R$ is injective on
$(-\varepsilon_1,\infty)$.  Moreover, $R$ is an injective immersion in
$(-\varepsilon_1,\infty)$, which is an open condition so that also
the second claim follows. Finally the assumption \textup{(iv)} with
$\mathcal{G}^{(0)}$ as in \textup{(\ref{Gzw-symm})} follows from
$0\neq z+w$ for all $z,w\in \mathcal{V}$, since
$\mathrm{Re}(z+w)=\mathrm{Re}(z) +\mathrm{Re}(w)>0$.  \hfill $\square$

\bigskip

We finish this section with the 
\\[\smallskipamount]
\noindent\emph{Proof of Theorem~\textup{\ref{thm:main}}.} 
We have seen that equation~(\ref{eq:GZW}) can be solved by
Ansatz~\ref{ansatz} to $G^{(0)}(R(z),R(w))=\mathcal{G}^{(0)}(z,w)$,
where $R$ is given in (\ref{eq2:R}) and $\mathcal{G}^{(0)}$ in
(\ref{calG:zhatw}) resp.\ \textup{(\ref{Gzw-symm})}.  This solution
depends on the choice of preimages $\varepsilon_k \in R^{-1}(E_k)$
made in (iv) of Ansatz~\ref{ansatz}. \emph{Any} solution
$\{\varepsilon_1,\dots, \varepsilon_d\}$ of the system of equations
(\ref{thm:R}) provides a solution of (\ref{calG:zhatw}), if also the
assumptions (iv), (v) and (vi) of Ansatz~\ref{ansatz} hold.
Theorem~\ref{thm:main} selects one particular solution of
(\ref{eq:GZW}) which satisfies the assumptions (ii), (iv), (v) and
(vi) of Ansatz~\ref{ansatz} by Lemma \ref{lemma:neighbourhood}. Hence
also relation (vii) of Ansatz~\ref{ansatz} holds by
Proposition~\ref{prop:vii}.  The choice $\lim_{\lambda\to 0}
\varepsilon_k=E_k$ and $\lim_{\lambda\to 0} \varrho_k=r_k$ is made to
recover in the limit $\lambda\to 0$ the moments of the Gau\ss{}ian
measure~(\ref{measure0}).

It remains to show (\ref{Gzw-new}). On the right hand side of
(\ref{GRR}) we use the the symmetry
$\mathcal{G}^{(0)}(\varepsilon_k,w)=
\mathcal{G}^{(0)}(w,\varepsilon_k)$ from Proposition~\ref{prp:symm}
and express $\mathcal{G}^{(0)}(w,\varepsilon_k)$ as (\ref{calG:zhatw})
for $w\mapsto \varepsilon_k$ and $z\mapsto w$.  Dividing by
$(R(w)-R(z))$ gives
\begin{align}
\mathcal{G}^{(0)}(z,w)=\frac{\displaystyle 
1 -\frac{\lambda}{N} \sum_{k=1}^d \frac{r_k}{
(R(z)-R(\varepsilon_k))(R(\varepsilon_k)-R({-}w))}
\prod_{j=1}^d \frac{
R(w){-}R({-}\widehat{\varepsilon_k}^j)}{ R(w)-R(\varepsilon_j)}
}{R(w)-R(-z)}\;.
\label{Gzw-final}
\end{align}
This equation was obtained in \cite{Grosse:2019jnv} by another
method. We rearrange it as 
\begin{align*}
\mathcal{G}^{(0)}(z,w)
&=\frac{1}{(R(w)-R(-z))(R(z)-R(-w))}
\Big\{
R(z)-R(-w) 
\\*
&-\frac{\lambda}{N} \sum_{l=1}^d \frac{r_l}{
(R(\varepsilon_l)-R({-}w))}
\prod_{j=1}^d \frac{
R(w)-R({-}\widehat{\varepsilon_l}^j)}{ R(w)-R(\varepsilon_j)}
\\
&-\frac{\lambda}{N} \sum_{k=1}^d \frac{r_k}{
(R(z)-R(\varepsilon_k))}
\prod_{j=1}^d \frac{
R(w)-R({-}\widehat{\varepsilon_k}^j)}{ R(w)-R(\varepsilon_j)}
\Big\}\;.
\end{align*}
The second line is $-\frac{\lambda}{N} \sum_{l=1}^d 
\mathcal{G}^{(0)}(w,\varepsilon_l)$ by (\ref{calG:zhatw}). We combine
it with the term $-R(-w)$ inside $\{~\}$ according to
our main algebraic relation (\ref{Rznegz}). In the last line, 
the factor $\prod_{j=1}^d \frac{
R(w)-R({-}\widehat{\varepsilon_k}^j)}{ R(w)-R(\varepsilon_j)}$ is 
rewitten via (\ref{identity}),  
with $w\mapsto \varepsilon_k$ and $z\mapsto w$. We arrive at
\begin{align*}
\mathcal{G}^{(0)}(z,w)
&=\frac{1}{(R(w)-R(-z))(R(z)-R(-w))}
\bigg\{
R(z)+R(w)
\\
&+\frac{\lambda}{N} \sum_{k=1}^d \frac{r_k}{
(R(\varepsilon_k)-R(z))}
+\frac{\lambda}{N} \sum_{l=1}^d \frac{r_l}{
(R(\varepsilon_l)-R(w))}
\\
&+\frac{\lambda}{N} \sum_{k,l=1}^d \frac{r_k}{
(R(\varepsilon_k)-R(z))(R(\varepsilon_l)-R(w))}
\frac{\prod_{j=1}^d 
(R(\varepsilon_l)-R({-}\widehat{\varepsilon_k}^j))}{ 
\prod_{j\neq l}^d (R(\varepsilon_l)-R(\varepsilon_j)}
\bigg\}\;.
\end{align*}
The result (\ref{Gzw-new}) follows from 
equation (\ref{calG:epshatw}) for 
$\mathcal{G}^{(0)}(\varepsilon_k,\varepsilon_l)$.
\hspace*{\fill}$\square$%

\section{The diagonal 2-point function}

The diagonal planar cumulant $z\mapsto \mathcal{G}^{(0)}(z,z)$ 
of length 2 and cycle type $(0,1)$ admits a simpler formula due to
properties of the rational function $\tilde{R}$ with 
$\tilde{R}(z):=R(z)-R(-z)$. 
Let $z\in \{0,\pm \alpha_1,\dots, \pm \alpha_d\}$ be the list of roots of 
\[
0= R(z)-R(-z)=2z -\frac{\lambda}{N} 
\sum_{k=1}^d \frac{\varrho_k}{\varepsilon_k+z} -\frac{\lambda}{N} 
\sum_{k=1}^d \frac{\varrho_k}{-\varepsilon_k+z} \:,
\] 
with the convention $\alpha_k>0$ and $\alpha_k\neq \alpha_l$ for $k\neq
l$. Since here all $\varepsilon_k>0$ and $\varrho_k>0$ are positive
real numbers, we can argue as for the rational function $R$ that also
the rational function $\tilde{R}$ maps
each of the $2d+1$ connected components of $\mathbb{R}\backslash
\{\pm\varepsilon_1, \dots, \pm \varepsilon_d\}$ bijectively onto
$\mathbb{R}$. So the odd function $\tilde{R}$ has indeed $2d+1$
different real roots $\{0,\pm \alpha_1,\dots, \pm \alpha_d\}$ of the
equation $\tilde{R}=0$. Taking its poles $\{\infty,\pm \varepsilon_k\}$ into
account, we have
\begin{align}
(R(z)-R(-z))
&= 2z \prod_{k=1}^d \frac{(z^2-\alpha_k^2)}{(z^2-\varepsilon_k^2)}\;.
\label{Rzz}
\end{align}
\begin{proposition}
\label{prop:Gww}
For any $z\in \hat{\mathbb{C}}$, the diagonal planar cumulant of cycle
type $(0,1)$ can be simplified to
\begin{align}
\mathcal{G}^{(0)}(z,z)&=
\frac{2 R(z)-2R(0)}{
(R(z)-R(-z))^2}\prod_{k=1}^d \frac{(R(z)-R(\alpha_k))^2}{
(R(z)-R(\varepsilon_k))^2}
\label{G2-diag}
\\
&\equiv \frac{1}{2z} \prod_{k=1}^d
\frac{(z-\hat{0}^k)(z+e_k)\prod_{j=2}^d (z-\widehat{\alpha_k}^j)^2}
{\prod_{l=1}^d (z-\widehat{e_k}^l)^2}\;.\nonumber
\end{align}
\end{proposition}
\noindent
\emph{Proof.} The $d+1$ fold product of (\ref{Rzz}) 
for $\{z,\hat{z}^1,\dots,\hat{z}^d\}$
is inserted into (\ref{calG:zhatw}):
\begin{align}
&(R(z)-R(-z))^2 \mathcal{G}^{(0)}(z,z) \prod_{k=1}^d  
(R(z)-R(\varepsilon_k))
\nonumber
\\[-2ex]
&=(R(z)-R(-z))
\prod_{l=1}^d(R(z)-R(-\hat{z}^l))
\nonumber
\\[-1ex]
&= 2z\Big(\prod_{l=1}^d 2\hat{z}^l\Big)
\prod_{k=1}^d \frac{(z^2-\alpha_k^2)\prod_{l=1}^d ((\hat{z}^l)^2-\alpha_k^2)
}{(z^2-\varepsilon_k^2)\prod_{l=1}^d ((\hat{z}^l)^2-\varepsilon_k^2)}\;.
\label{Gzz-1}
\end{align}
We use cases of (\ref{R-rational}); the third one 
takes $R(\alpha_k)=R(-\alpha_k)$ into account:\enlargethispage{5mm}
\begin{align*}
2(R(z)-R(0))&=\frac{2z\prod_{l=1}^d (2\hat{z}_l) }{
\prod_{k=1}^d (-2\varepsilon_k)}\;,
\\[-0.5ex]
(R(z)-R(\varepsilon_k))
&=(z-\varepsilon_k)\frac{\prod_{l=1}^d (\varepsilon_k-\hat{z}^l) }{
\prod_{l=1}^d (\varepsilon_k+\varepsilon_l)}\;,
\\
(R(z)-R(\alpha_k))^2&=(z^2-\alpha_k^2)
\frac{\prod_{l=1}^d ((\hat{z}^l)^2-\alpha_k^2) }{
\prod_{l=1}^d (\varepsilon_l^2-\alpha_k^2)}\;.
\end{align*}
We identify in (\ref{Gzz-1}) the first equation and 
the product over $k$ of the second and third equations:
\begin{align*}
&(R(z)-R(-z))^2 \mathcal{G}^{(0)}(z,z) \prod_{k=1}^d  
(R(z)-R(\varepsilon_k))
\nonumber
\\*[-1ex]
&= \frac{2(R(z)-R(0))}{
\prod_{k=1}^d ((z+\varepsilon_k)
\prod_{l=1}^d (\hat{z}^l+\varepsilon_k))}
\prod_{k=1}^d\frac{(R(z)-R(\alpha_k))^2}{
(R(z)-R(\varepsilon_k))} 
\cdot \prod_{k=1}^d  \frac{(2\varepsilon_k)
\prod_{l=1}^d (\varepsilon_l^2-\alpha_k^2)}{
\prod_{l=1}^d (\varepsilon_l+\varepsilon_k)}\;.
\end{align*}
Now observe that the residue of (\ref{R-rational}) at
$z=-\varepsilon_k$ is the identity
\begin{align}
\frac{(u+\varepsilon_k)\prod_{l=1}^d 
(\hat{u}^l+\varepsilon_k)}{\prod_{j\neq k}^d 
(\varepsilon_j-\varepsilon_k)} = -\frac{\lambda}{N} \varrho_k\;,
\end{align}
for any $u\notin R^{-1}(\{\infty\})$. Consequently, 
\begin{align*}
\frac{(R(z)-R(-z))^2 \mathcal{G}^{(0)}(z,z)}{2(R(z)-R(0))} 
\prod_{k=1}^d\frac{(R(z)-R(\varepsilon_k))^2}{(R(z)-R(\alpha_k))^2}
&= \prod_{k=1}^d  \frac{ N
\prod_{l=1}^d (\varepsilon_l^2-\alpha_k^2)}{(-\lambda\varrho_k)\prod_{j\neq k}^d 
(\varepsilon_j^2-\varepsilon_k^2)}=C
\end{align*}
is a constant independent of $z$, which for $z\to \infty$ is identified as
$C=1$. \hspace*{\fill} $\square$%

\bigskip

The following result will be needed in the next section:
\begin{lemma}\label{lemma2}
For any $w\in \hat{\mathbb{C}}$ one has 
\begin{align}
\frac{1}{R(w)-R({-}w)}
+ \sum_{k=1}^d 
\frac{1}{R(w)-R({-}\hat{w}^k)}
&=
\frac{1}{2(R(w){-}R(0))}
+\sum_{k=1}^d \frac{1}{R(w)-R(\alpha_k)}\;.
\label{fractions}
\end{align}
\end{lemma}
\noindent\emph{Proof.} Taking
$R(w)=R(\hat{w}^k)$ into account, 
all terms on the lhs of (\ref{fractions}) are of the form (\ref{Rzz}) so that 
the lhs of
(\ref{fractions}) has simple poles at
$w \in \{0,\pm \alpha_l\}$ and $\hat{w}^k\in \{0,\pm
\alpha_l\}$. Applying $R$ shows that these $\hat{w}^k$ correspond to
additional poles at $w \in \{\hat{0}^l, \widehat{\alpha_l}^j,\mp
\alpha_l\}$ for
$l=1,\dots,d$ and $j=2,\dots,d$. We evaluate the residues at these
poles and check that the rhs of (\ref{fractions}) 
has the same poles (clear) with the same residues.

Note that $R(w)=R(\hat{w}^k)$ implies 
$R'(w) =R'(\hat{w}^k) (\hat{w}^k)'(w)$ or 
$(\hat{w}^k)'(w)=\frac{R'(w)}{R'(\hat{w}^k)}$. 
Consider the pole at $w=\pm \alpha_l$. Then there is precisely one $k_l 
\in \{1,\dots, d\}$ with $\hat{w}^{k_l}=\mp \alpha_l$. Therefore,
\begin{align*}
&\Res\displaylimits_{w= \pm \alpha_l}
\Big(\frac{dw }{R(w)-R(-w)}
+ \sum_{k=1}^d 
\frac{dw}{R(w)-R(-\hat{w}^k)}
\Big)
\nonumber
\\
&= \Big(\frac{1}{R'(w)+R'(-w)}+ 
\frac{1}{R'(w)+R'(-\hat{w}^{k_l})
\frac{R'(w)}{R'(\hat{w}^{k_l})}}\Big)
\Big|_{w=\pm \alpha_l,\hat{w}^{k_l}=\mp\alpha_l}
=\frac{1}{R'(\pm \alpha_l)}\;.
\end{align*}
Consider in case of $d\geq 2$ the pole at $w=\widehat{\alpha_l}^j$. 
There are precisely two distinct $k_+,k_-
 \in \{1,\dots, d\}$ with 
 $\hat{w}^{k_+}=\alpha_l$ and 
 $\hat{w}^{k_-}=-\alpha_l$. 
 Therefore,
\begin{align*}
&\Res\displaylimits_{w= \widehat{\alpha_l}^j}
\Big(\frac{dw }{R(w)-R(-w)}
+ \sum_{k=1}^d 
\frac{dw}{R(w)-R(-\hat{w}^k)}
\Big)
\nonumber
\\
&= 
\Big(\frac{1}{R'(w)+R'(-\hat{w}^{k_+})
\frac{R'(w)}{R'(\hat{w}^{k_+})}}
+\frac{1}{R'(w)+R'(-\hat{w}^{k_-})
\frac{R'(w)}{R'(\hat{w}^{k_-})}}\Big)
\Big|_{w=\widehat{\alpha_l}^j,\hat{w}^{k_\pm}=\pm\alpha_l}
\\
&=\frac{1}{R'(\widehat{\alpha_l}^j)}\;.
\end{align*}
The rhs of (\ref{fractions}) has exactly the same residues. 

Finally, it is also clear that both sides of (\ref{fractions}) 
have the same residue $\frac{1}{2R'(0)}$ 
at $w=0$. 
For $w=\hat{0}^l$ there is a unique $k_l \in \{1,\dots, d\}$ with 
$\hat{w}^{k_l}=0$. Then 

\begin{align*}
&\Res\displaylimits_{w= \widehat{0}^l}
\Big(\frac{dw }{R(w)-R(-w)}
+ \sum_{k=1}^d 
\frac{dw}{R(w)-R(-\hat{w}^k)}
\Big)
\nonumber
\\
&= \frac{1}{R'(w)+R'(-\hat{w}^{k_l})
\frac{R'(w)}{R'(\hat{w}^{k_l})}}
\Big|_{w=\hat{0}^l,\hat{w}^{k_l}=0}
=\frac{1}{2R'(\hat{0}^l)}\;,
\end{align*}
which agrees with the residue of the rhs
of (\ref{fractions}). Therefore,
the difference between lhs and rhs of  (\ref{fractions}) is a bounded 
entire function, i.e.\ a constant by Liouville's theorem, which is 
zero when considering $w\to \infty$. This finishes the proof. 
\hspace*{\fill} $\square$%

\section{The planar $1+1$-point function}

The $1+1$-point function $G_{|a|b|}$ is the cumulant of length 2 and cycle
type $(2,0)$ (i.e.\ two cycles $a$ and $b$ of length $1$), see
Appendix \ref{appA}. We derive in Appendix~\ref{appB} its equation of
motion (\ref{DS11-final}) whose restriction to the planar sector (of
degree or genus $g=0$) reads
\begin{align}
(E_a+E_a)G_{|a|b|}^{(0)} &=
-\frac{\lambda}{N} 
\sum_{k=1}^{N} G^{(0)}_{|ak|}G^{(0)}_{|a|b|}
+\frac{\lambda}{N} 
\sum_{k=1}^{N}\frac{G^{(0)}_{|k|b|}-G^{(0)}_{|a|b|}}{E_k-E_a}
+\lambda
\frac{G^{(0)}_{|bb|}-G^{(0)}_{|ab|}}{E_b-E_a}\;.
\label{G11}
\end{align}
We interpret this equation as evaluation 
$G_{|a|b|}^{(0)}=\mathcal{G}^{(0)}(\varepsilon_a|\varepsilon_b)$ of a 
function\footnote{Be careful to distinguish
$\mathcal{G}^{(0)}(z|w)$ from $\mathcal{G}^{(0)}(z,w)$.}  
$\mathcal{G}^{(0)}(z|w)$ which satisfies 
\begin{align}
\big( R(z)-R(-z)\big) \mathcal{G}^{(0)}(z|w)
- \frac{\lambda}{N} \sum_{k=1}^d
\frac{r_k\mathcal{G}^{(0)}(\varepsilon_k|w) }{
R(\varepsilon_k)-R(z)}
&=
\lambda \frac{\mathcal{G}^{(0)}(z,w)-\mathcal{G}^{(0)}(w,w)}{
R(z)-R(w)}\;.
\label{calG11:GW}
\end{align}
The identity (\ref{Rznegz}) was decisive here, and 
multiplicities $r_k$ of the $E_k=R(\varepsilon_k)$
were admitted.

Since  
$\mathcal{G}^{(0)}(\alpha_k|w)$ must be regular\footnote{Regularity
of $\mathcal{G}^{(0)}(\alpha_k|w)$ is here a technical assumption
which is justified by viewing (\ref{calG11:GW}) as limiting case of 
singular integral equations of Carleman type (see
e.g.\ \cite[\S 4.4]{Tricomi:1957??}).
Their solutions are regular for any $z,w>0$.} for any $w>0$,
evaluation at $z=\alpha_k$ produces $d$ equations 
\begin{align}
\frac{\lambda}{N} \sum_{l=1}^d
\frac{r_l\mathcal{G}^{(0)}(\varepsilon_l|w) }{
R(\alpha_k)-R(\varepsilon_l)}
&=
\lambda \frac{\mathcal{G}^{(0)}(\alpha_k,w)-\mathcal{G}^{(0)}(w,w)}{
R(\alpha_k)-R(w)}\;.
\label{calG11-alph}
\end{align}
Equation (\ref{calG11-alph}) is with 
Proposition~\ref{prop:Schechter} solved by
\begin{align*}
\frac{r_k}{N}  \mathcal{G}^{(0)}(\varepsilon_k|w) 
= 
\sum_{l=1}^{d} 
(R(\alpha_l)-R(\varepsilon_k)) \mathbf{A}_l(R(\varepsilon_k))
\mathbf{E}_k(R(\alpha_l)) 
\frac{\mathcal{G}^{(0)}(\alpha_l,w)-\mathcal{G}^{(0)}(w,w)}{
R(\alpha_l)-R(w)}\;,
\end{align*}
where 
\begin{align*}
\mathbf{A}_i(x)
&=\frac{\mathbf{A}(x)}{(x-R(\alpha_i))\mathbf{A}'(R(\alpha_i))} 
&
\text{and} \quad \mathbf{E}_j(y)&=\frac{\mathbf{E}(y)}{(y-R(\varepsilon_j))
\mathbf{E}'(R(\varepsilon_j))}\:,
\\
\text{with}\quad
\mathbf{A}(x)&:=\prod_{k=1}^d (x-R(\alpha_k)) & \text{and} \quad 
\mathbf{E}(y)&=\prod_{k=1}^d (y-R(\varepsilon_k)) \:.
\end{align*}
Here, the $R(\alpha_k)$ and
$R(\varepsilon_l)$ are pairwise distinct, since $R$ is injective on
$(-\varepsilon_1,\infty)$. Inserting this back into 
(\ref{calG11:GW}) gives the $1+1$-point function
\begin{align}
\mathcal{G}^{(0)}(z|w)
&= \frac{\lambda}{R(z)-R(-z)}
\Big\{
\frac{\mathcal{G}^{(0)}(z,w)-\mathcal{G}^{(0)}(w,w)}{
R(z)-R(w)}
\nonumber
\\
&-\sum_{k,l=1}^{d}
\frac{(R(\alpha_l)-R(\varepsilon_k)) \mathbf{A}_l(R(\varepsilon_k))
\mathbf{E}_k(R(\alpha_l)) }{R(z)-R(\varepsilon_k)}
\frac{\mathcal{G}^{(0)}(\alpha_l,w)-\mathcal{G}^{(0)}(w,w)}{
R(\alpha_l)-R(w)}
\Big\}
\label{SW31}
\end{align}
in terms of the 2-point function $\mathcal{G}^{(0)}(z,w)$ 
known from Theorem~\ref{thm:main}.
We convert the solution (\ref{SW31}) into a manifestly symmetric form: 
\begin{proposition}
\label{prop:G11-symm}
The planar cumulant of length $2$ and cycle type $(2,0)$ has the solution
\begin{align}
\mathcal{G}^{(0)}(z|w)
&= \frac{\lambda}{(R(z)-R(w))^2} \Big(
\mathcal{G}^{(0)}(z,w)
\label{G11-symm}
\\
&-
\frac{R(z)+R(w)-2R(0)}{
(R(z){-}R({-}z))(R(w){-}R({-}w))}
\prod_{k=1}^d \frac{(R(z)-R(\alpha_k))(R(w)-R(\alpha_k))}{
(R(z)-R(\varepsilon_k))(R(w)-R(\varepsilon_k))}\Big)\,.
\nonumber
\end{align}
\end{proposition}
\noindent\emph{Proof.} Using (\ref{Schechter12}) 
we evaluate the $k$-sum in (\ref{SW31}) to
\begin{align}
\mathcal{G}^{(0)}(z|w)
&= \frac{\lambda}{R(z)-R(-z)}
\Big\{
\frac{\mathcal{G}^{(0)}(z,w)-\mathcal{G}^{(0)}(w,w)}{
R(z)-R(w)}
\nonumber
\\
&-\sum_{l=1}^{d} 
\frac{\mathbf{A}_l(R(z))\mathbf{E}(R(\alpha_l))}{
\mathbf{E}(R(z))}
\frac{\mathcal{G}^{(0)}(\alpha_l,w)-\mathcal{G}^{(0)}(w,w)}{
R(\alpha_l)-R(w)}
\Big\}\;.
\label{G11-sumk}
\end{align}
In the second line we have 
\[
\frac{\mathbf{A}_l(R(z))\mathbf{E}(R(\alpha_l))}{
\mathbf{E}(R(z))}= 
-\frac{\mathbf{A}(R(z))}{\mathbf{E}(R(z))}
\cdot \frac{
\prod_{k=1}^d (R(\alpha_l)-R(\varepsilon_k))}{
(R(\alpha_l)-R(z))
\prod_{j=1,j\neq l}^d (R(\alpha_l)-R(\alpha_j))}\;,
\]
and we recall
\[
\mathcal{G}^{(0)}(\alpha_l,w)=-\frac{1}{(R(\alpha_l)-R(w))} 
\frac{\prod_{j=1}^d R(\alpha_l)-R(-\hat{w}^j)}{
\prod_{j=1}^d (R(\alpha_l)-R(\varepsilon_j))}
\]
from (\ref{calG:zhatw}). Inserting both identities into
(\ref{G11-sumk}) gives after a first partial fraction decomposition
\begin{align}
\mathcal{G}^{(0)}(z|w)
&= \frac{\lambda}{(R(z)-R(-z))(R(z)-R(w))}
\Big\{
\mathcal{G}^{(0)}(z,w)-\mathcal{G}^{(0)}(w,w)
\nonumber
\\
&-\frac{\mathbf{A}(R(z))}{\mathbf{E}(R(z))}\mathcal{G}^{(0)}(w,w)
\Big(
\sum_{l=1}^{d} 
\frac{
\prod_{k=1}^d (R(\alpha_l)-R(\varepsilon_k))}{
(R(\alpha_l)-R(z))\prod_{j=1,j\neq l}^d (R(\alpha_l)-R(\alpha_j))}
\nonumber
\\*
&\qquad\qquad\qquad
-\sum_{l=1}^{d} \frac{
\prod_{k=1}^d (R(\alpha_l)-R(\varepsilon_k))}{
(R(\alpha_l)-R(w))\prod_{j=1,j\neq l}^d (R(\alpha_l)-R(\alpha_j))}
\Big)
\nonumber
\\
&-\frac{\mathbf{A}(R(z))}{\mathbf{E}(R(z))}\Big(
\sum_{l=1}^{d} 
\frac{
\prod_{k=1}^d (R(\alpha_l)-R({-}\hat{w}^k))}{
(R(\alpha_l){-}R(z))(R(\alpha_l){-}R(w))
\prod_{j=1,j\neq l}^d (R(\alpha_l){-}R(\alpha_j))}
\nonumber
\\*
&\qquad\qquad\qquad
-\sum_{l=1}^{d} 
\frac{
\prod_{k=1}^d (R(\alpha_l)-R(-\hat{w}^k))}{
(R(\alpha_l){-}R(w))^2\prod_{j=1,j\neq l}^d (R(\alpha_l){-}R(\alpha_j))}\Big)
\Big\}\,.
\label{Gzw-2}
\end{align}
The second and third line are converted via an identity 
(\ref{identity}) with substitution
$\varepsilon_i\mapsto \alpha_i$ and $-\hat{w}^j\mapsto
\varepsilon_j$. One of the surviving terms cancels 
$\mathcal{G}^{(0)}(w,w)$ in the first line of (\ref{Gzw-2}). 
Another partial fraction decomposition in the fourth line of
(\ref{Gzw-2})
and 
$\frac{1}{(R(\alpha_l)-R(w))^2}
=\lim_{u\to w} 
\frac{1}{(R(u)-R(w))}\big(
\frac{1}{(R(\alpha_l)-R(u))}-
\frac{1}{(R(\alpha_l)-R(w))}\big)$
in the fifth line of (\ref{Gzw-2}) also give rise to expressions 
(\ref{identity}) with substitution
$\varepsilon_i\mapsto \alpha_i$. We thus find
\begin{align*}
\mathcal{G}^{(0)}(z|w)
&\!=\! \frac{\lambda}{(R(z){-}R({-}z))(R(z){-}R(w))}\!
\bigg\{
\mathcal{G}^{(0)}\:\!\!(z,w){-}
\frac{\mathbf{A}(R(z))\mathbf{E}(R(w))}
{\mathbf{E}(R(z))\mathbf{A}(R(w))}\mathcal{G}^{(0)}\:\!\!(w,w)
\nonumber
\\
&+\frac{\displaystyle
\prod_{k=1}^{d} 
\frac{R(z)-R(-\hat{w}^k)}{R(z)-R(\varepsilon_k)}
-
\frac{\mathbf{A}(R(z))\mathbf{E}(R(w))}{\mathbf{E}(R(z))\mathbf{A}(R(w))}
\prod_{k=1}^{d} 
\frac{R(w)-R(-\hat{w}^k)}{R(w)-R(\varepsilon_k)}
}{R(z)-R(w)} 
\nonumber
\\
&-\frac{\mathbf{A}(R(z))}{\mathbf{E}(R(z))}
\lim_{u\to w} 
\frac{\displaystyle 
\prod_{k=1}^{d} 
\frac{R(u)-R(-\hat{w}^k)}{R(u)-R(\alpha_k)}
-
\prod_{k=1}^{d} 
\frac{R(w)-R(-\hat{w}^k)}{R(w)-R(\alpha_k)}
}{R(u)-R(w)}
\bigg\}\;.
\end{align*}
After evaluation of the limit we reconstruct in the last two lines 
$\mathcal{G}^{(0)}(z,w)$ and $\mathcal{G}^{(0)}(w,w)$ 
via (\ref{calG:zhatw}):
\begin{align*}
\mathcal{G}^{(0)}(z|w)
&= \frac{\lambda}{(R(z)-R(w))^2} \Big(
\mathcal{G}^{(0)}(z,w)
\nonumber
\\
&-
\frac{\mathbf{A}(R(z))\mathbf{E}(R(w))}{\mathbf{E}(R(z))\mathbf{A}(R(w))}
\frac{R(w)-R(-w)}{R(z)-R(-z)}
\mathcal{G}^{(0)}(w,w)
\Big\{1 \nonumber
\\
&+
\frac{R(z)-R(w)}{R(w)-R(-w)}
+\sum_{k=1}^d \frac{R(z)-R(w)}{R(w)-R(-\hat{w}^k)}
- \sum_{k=1}^d 
\frac{R(z)-R(w)}{R(w)-R(\alpha_k)}
\Big\}\Big)\;.
\end{align*}
With Lemma~\ref{lemma2} the terms in $\{~\}$ 
can be reduced to
$
\{~\}= \frac{R(z)+R(w)-2R(0)}{2(R(w)-R(0))}
$. Inserting (\ref{G2-diag}) for 
$\mathcal{G}^{(0)}(w,w)$
gives the final result (\ref{G11-symm}).
\hspace*{\fill}$\square$% 

\section{Outlook}

\label{sec:outlook}

We have developed a new algebraic solution strategy for
the two initial cumulants of a quartic analogue of the Kontsevich
model. Our results have been extended in \cite{Branahl:2020yru} to an
algorithm which allows to recursively compute all other cumulants. The
key discovery of \cite{Branahl:2020yru} was to understand that one
first has to focus on three families $\Omega^{(g)}_{m}(u_1,...,u_m)$,
$\mathcal{T}^{(g)}(u_1,...,u_m\|z,w|)$ and
$\mathcal{T}^{(g)}(u_1,...,u_m\|z|w|)$ of auxiliary functions. 
Their simplest cases are the functions 
$\mathcal{T}^{(0)}(\emptyset \|z,w|):=\mathcal{G}^{(0)}(z,w)$ and
$\mathcal{T}^{(0)}(\emptyset \|z|w|):=\mathcal{G}^{(0)}(z|w)$
analysed in this paper.  The
auxiliary functions are special polynomials \cite{Branahl:2020uxs} in
the original cumulants. One first solves a coupĺed system of equations
for $(\Omega^{(g)}_{m},\mathcal{T}^{(g)})$ and then uses the 
result to turn the Dyson-Schwinger equations for the cumulants into 
a problem which can easily be solved by inversion of Cauchy matrices.

Of particular interest are the functions $\Omega^{(g)}_{n}$
which give rise to a family of meromorphic differentials 
\begin{align}
\omega_{g,n}(z_1,...,z_n)=\Omega^{(g)}_{n}(z_1,...,z_n) 
dR(z_1)\cdots dR(z_n)
\end{align}
which starts with
$\omega_{0,2}(z_1,z_2)=\frac{dz_1\,dz_2}{(z_1-z_2)^2}
+\frac{dz_1\,dz_2}{(z_1+z_2)^2}$. Also the next forms $\omega_{0,3}$,
$\omega_{0,4}$ and $\omega_{1,1}$ have been found in
\cite{Branahl:2020yru}, where $\omega_{1,1}$ needs 
Propositions~\ref{prop:Gww} and \ref{prop:G11-symm} of this paper.
Remarkably, all forms computed so far
satisfy abstract loop equations \cite{Borot:2013lpa} if one sets
$\omega_{0,1}(z)=y(z)dx(z)$ with $x(z)=R(z)$ and $y(z)=-R(-z)$. It was
shown in \cite{Borot:2015hna} that the solution of abstract loop
equations is \emph{blobbed topological recursion}, a systematic
extension of topological recursion \cite{Eynard:2007kz,
  Eynard:2016yaa} by additional terms which are holomorphic at
ramification points of $x$. The natural conjecture is that all
$\omega_{g,n}$ of the quartic analogue of the Kontsevich model obey
blobbed topological recursion. The conjecture was proved for genus
$g=0$ in \cite{Hock:2021tbl} by relating it to an equation which
expresses 
$\omega_{g,n+1}(z_1,...,z_n,-z)$ in terms of 
$\omega_{g,m+1}(z_1,...,z_m,+z)$ with $m\leq n$.

In an early version of this paper we had speculated that the exact
solution of the non-linear equation~(\ref{eq:GZW}) might be caused by a
hidden integrable structure. The discovery in
\cite{Branahl:2020yru,Hock:2021tbl} that the quartic analogue of
the Konsevich model obeys blobbed topological recursion questions this
interpretation: integrability is not known in blobbed topological
recursion.  The relation to intersection theory on the moduli space
$\overline{\mathcal{M}}_{g,n}$ of stable complex curves extends,
however, to blobbed topological recursion \cite{Borot:2015hna}. The
discovery in \cite{Hock:2021tbl} that (at least the planar sector
of) the quartic analogue of the Kontsevich model is completely
governed by the behaviour of the $\omega_{g,n}$ under a global (and
canonical) involution makes us confident that the intersection numbers
generated by this model will have a geometric significance. It will
be an exciting programme to make this precise.

\appendix

\section{Decomposition of moments via cumulants}

\label{appA}

The moments (\ref{moments}) decompose into cumulants 
(see e.g.\ \cite{McCullagh2012,Speed}),
\begin{align}\label{cum}
\Big\langle \prod_{i=1}^n e_{k_il_i}\Big\rangle 
=\sum_{\substack{\text{partitions} \\ \text{$\pi$ of 
$\{1,\dots, n\}$}}} \prod_{\text{blocks $\beta \in \pi$}} 
\Big\langle \prod_{i\in \beta}  e_{k_i l_i} \Big\rangle_c\;.
\end{align}
There is a similar formula expressing the cumulants in terms of moments \cite[Eq.(1.2)]{Speed},
related to (\ref{cum}) via \emph{M\"{o}bius inversion} on the partially ordered set of (partitions of)  subsets of indices 
(the \emph{partition lattice} of $[N]\times [N]$)
$$I:=\{k_1l_1,\dots,k_nl_n\}\subset [N]\times [N]\:,$$
with $|I|=n$ and $[N]:=\{1,\dots,N\}$.
For a quartic potential (\ref{measure4}), moments and cumulants are only
non-zero if $n$ is even and every block $\beta$ is of even length. For
example,
\begin{align*}
\langle e_{k_1l_1}e_{k_2l_2}e_{k_3l_3}e_{k_4l_4}
\rangle 
&=\langle e_{k_1l_1}e_{k_2l_2}e_{k_3l_3}e_{k_4l_4}
\rangle_c + 
\langle e_{k_1l_1}e_{k_2l_2}\rangle_c \langle e_{k_3l_3}e_{k_4l_4}\rangle_c 
\\
&+\langle e_{k_1l_1}e_{k_3l_3}\rangle_c \langle e_{k_2l_2}e_{k_4l_4}\rangle_c 
+\langle e_{k_1l_1}e_{k_4l_4}\rangle_c \langle e_{k_2l_2}e_{k_3l_3}\rangle_c 
\;.
\end{align*}

Note that in our context the moments $\big\langle \prod_{i=1}^n
e_{k_il_i}\big\rangle$ are invariant under permutations of
$I:=\{k_1l_1,\dots,k_nl_n\}$ so that they only depend on the subset
$I\subset [N]\times [N]$, but not on the choice of a labelling
$I=\{k_1l_1,\dots,k_nl_n\}\simeq [n]$. By \cite[Eq.(1.2)]{Speed} the
same is then true for the cumulants, i.e.\ $\big\langle \prod_{i\in
  \beta} e_{k_i l_i} \big\rangle_c$ only depends on the subset
$\{k_il_i|\: i\in \beta\}\subset [N]\times [N]$.  

We restrict our attention to the case that all $k_i$ are pairwise
different.  Then the structure of the Gau\ss{}ian measure
(\ref{measure0}) (together with the invariance of a trace under cyclic
permutations) implies that the cumulant
$\big\langle \prod_{i=1}^n e_{k_il_i}\big\rangle_c$ corresponding to
$I=\{k_1l_1,\dots,k_nl_n\}$ with $|I|=n>0$ is only non-zero if $I$ has
a permutation $\sigma$ with $pr_2=pr_1\circ \sigma$. Here
$$pr_i: [N]\times [N]\supset I\to [N]$$ is the projection onto the
corresponding factor for $i=1,2$. By choosing a
labelling $$I:=\{k_1l_1,\dots,k_nl_n\}\simeq [n]$$ as before, this
corresponds to a permutation $\sigma$ in the symmetric group
$ \mathcal{S}_n$, with
$(l_1,\dots,l_n)=(k_{\sigma(1)},\dots,k_{\sigma(n)})$.

Therefore, the
cumulant $\big\langle \prod_{i=1}^n e_{k_il_i}\big\rangle_c$ only
depends on $I$ and the \emph{conjugacy class} of a permutation in $
\mathcal{S}_n$ (corresponding to the permutation $\sigma$ of $I$ with
$pr_2=pr_1\circ \sigma$), which is again independent of the choice of
the labelling of $I$.  In fact such conjugacy classes in $
\mathcal{S}_n$ just correspond to the different \emph{cycle types} of
a permutation in the symmetric group $ \mathcal{S}_n$. 
The cycle type of $\sigma$ is the $n$-tuple
$(\ell_1(\sigma),\ell_2(\sigma),\dots,\ell_n(\sigma))$ where
$\ell_k(\sigma)$ is the number of cycles of length $k$ in $\sigma$, 
with $\sum_{i=1}^n i\ell_i(\sigma)=n$. 
The number of cycles in a permutation $\sigma$ is 
$b(\sigma)=\sum_{i=1}^n \ell_i(\sigma)$. The number of 
different cycle types is the partition number $p(n)$, and there are  
$\frac{n!}{
1^{\ell_1} \ell_1! 2^{\ell_2}\ell_2!\dots 
n^{\ell_n} \ell_n!}$ permutations with the same cycle type 
$(\ell_1,\dots,\ell_n)$. 

Conversely, the $l$-indices of a 
non-vanishing cumulant $\langle e_{k_1l_1}\dots e_{k_nl_n}\rangle_c$ 
are completely determined by the cycle type and the information which 
$k$'s belong in which cyclic order to the same cycle. If, after 
renaming the $k$'s, $(k_1^1,\dots,k^1_{n_1})$ belong 
to one cycle,  $(k_1^2,\dots,k^2_{n_2})$ belong 
to another cycle, and so on up to the $b^{\text{th}}$ cycle, 
this information uniquely encodes a cumulant (with $n=n_1+\dots+n_b$)
\begin{align}
N^n \big\langle (e_{k_1^1k_2^1} 
e_{k_2^1k_3^1} \cdots 
e_{k_{n_1}^1k_1^1}) \cdots 
(e_{k_1^bk_2^b} e_{k_2^bk_3^b} \cdots 
e_{k_{n_b}^bk_1^b}) \big\rangle_c
=: 
N^{2-b} G_{|k_1^1\dots k_{n_1}^1|\dots
|k_1^b\dots k_{n_b}^b|} \;.
\label{cumulants-cycletype}
\end{align}

The power series expansion of the 
Fourier transform $\mathcal{Z}(M)$ into moments (\ref{moments}) 
can be compared with the 
insertion of (\ref{cumulants-cycletype}) into (\ref{cum}). 
The first terms are:
\begin{align}
\mathcal{Z}(M)&= 
1- \frac{1}{N^2} \sum_{j,k=1}^N \Big\{\frac{N}{2}
G_{|jk|} M_{jk}M_{kj}
+ \frac{1}{2} G_{|j|k|} M_{jj}M_{kk}\Big\}
\nonumber
\\
&
+ \frac{1}{N^4}\sum_{j,k,l,m=1}^N \Big\{
\frac{N}{4}G_{|jklm|} M_{jk} M_{kl}M_{lm}M_{mj}
+ \frac{1}{3} G_{|j|klm|} M_{jj}M_{kl} M_{lm}M_{mk}
\nonumber
\\
&+ \frac{1}{8} G_{|jk|lm|} M_{jk}M_{kj}M_{lm}M_{ml}
+ \frac{1}{4N}G_{|j|k|lm|} M_{jj}M_{kk} M_{lm}M_{ml}
\nonumber
\\
& 
+ \frac{1}{24N^2} G_{|j|k|l|m|} M_{jj}M_{kk}M_{ll}M_{mm}
+ \frac{N^2}{8} G_{|jk|} G_{|lm|} M_{jk}M_{kj} M_{lm}M_{ml}
\nonumber
\\
& + \frac{N}{4} G_{|jk|} G_{|l|m|} M_{jk}M_{kj} M_{ll}M_{mm}
+ \frac{1}{8} G_{|j|k|} G_{|l|m|} 
M_{jj} M_{kk}M_{ll} M_{mm}\Big\}
\nonumber
\\
&+ \mathcal{O}(M^6)\;.
\label{calZ4}
\end{align}

\section{Equations for the second cumulant}

\label{appB}

We derive here equations for the two non-vanishing second-order
cumulants $G_{|ab|} =\frac{1}{N}\langle e_{ab} e_{ba}\rangle_c$ of
cycle type $(0,1)$ (i.e.\ one cycle $ab$ of length $2$) and $G_{|a|b|}
=\langle e_{aa} e_{bb}\rangle_c$ of cycle type $(2,0)$ (i.e.\ two
cycles $a$ and $b$ of length $1$). To distinguish
$G_{|ab|}$ and $G_{|a|b|}$ we require $a\neq b$.

We start from (\ref{eom}) with $\mathcal{Z}(M)$ given by
(\ref{calZ4}), apply $\frac{N(E_a+E_b)}{\mathrm{i}}
\frac{\partial}{\partial M_{ba}}$ and put $M=0$. For $a\neq b$ this
gives the following result (the underlining should be ignored for the
moment; we explain it later):
\begin{align}
(E_a+E_b)G_{|ab|}&=1- 
\lambda
\Big\{ 
\frac{1}{N^2}
\sum_{k,l=1}^{N} \underline{\underline{G_{|bakl|}}}
+ \frac{1}{N}
\sum_{k=1}^{N} (G_{|ab|}G_{|ak|}+\underline{\underline{G_{|ab|} 
G_{|bk|}}})
\nonumber
\\[-1ex]
&+\frac{1}{N^2}
\big(
G_{|abab|}+\underline{G_{|abbb|}} +G_{|baaa|}
+G_{|ab|}(G_{|a|a|}+\underline{G_{|a|b|}}+\underline{G_{|b|b|}})\big)
\nonumber
\\
&+\frac{1}{N^3} \sum_{k=1}^{N} (
\underline{\underline{G_{|k|bak|}}}
+\underline{\underline{G_{|a|bak|}}}+ \underline{G_{|b|bak|}}+
\underline{\underline{G_{|ab|bk|}}}+G_{|ab|ak|})
\nonumber
\\
&+\frac{1}{N^4} (G_{|a|a|ab|}+\underline{G_{|a|b|ab|}}
+\underline{G_{|b|b|ab|}})\Big\}\;.
\label{DS2}
\end{align}

Next, we set $b \equiv a$ in (\ref{eom}) for 
$\mathcal{Z}(M)$ given by (\ref{calZ4}), apply
$\frac{N^2(E_a+E_a)}{\mathrm{i}} 
\frac{\partial}{\partial M_{bb}}$ for $a \neq 
b$ and obtain for $M=0$ (ignore again the underlining):
\begin{align}
(E_a+E_a)G_{|a|b|}&=-\lambda 
\Big\{ 
\underline{\underline{G_{|bb|}G_{|ab|}}}
+\frac{1}{N^2}
\sum_{k,l=1}^{N} \underline{G_{|b|akl|}}
\nonumber
\\[-1ex]
&+\frac{1}{N}
\sum_{k=1}^{N} (\underline{G_{|bbka|}}+\underline{\underline{G_{|bbak|}}}
+G_{|ak|}G_{|a|b|}
+\underline{G_{|ak|}G_{|a|b|}}
+\underline{G_{|ak|}G_{|b|k|}})
\nonumber
\\
&+\frac{1}{N^2} 
(G_{|b|aaa|}+G_{|a|abb|}+
\underline{\underline{G_{|a|abb|}}}+\underline{\underline{G_{|b|abb|}}}
+\underline{\underline{G_{|bb|ab|}}}+3 G_{|a|b|}G_{|a|a|})
\nonumber
\\
&+\frac{1}{N^3}
\sum_{k=1}^{N} (
G_{|a|b|ak|}+\underline{G_{|a|b|ak|}}
+\underline{G_{|b|k|ak|}})
+\frac{1}{N^4} G_{|b|a|a|a|}\Big\}\;.
\label{DS11}
\end{align}

Equations (\ref{DS2}) and (\ref{DS11}) are the analogues of
Dyson-Schwinger equations in quantum field theory. In this form they
provide little information because the right hand sides are too
complicated. We will now establish from the equations of motion
(\ref{eom}) two other identities which collect the underlined terms in
(\ref{DS2}) and (\ref{DS11}) into a function of the left hand sides.

To establish the identities, set $b\mapsto k$ in (\ref{eom}) and apply
$\frac{N(E_a+E_k)}{\mathrm{i}} \frac{\partial}{\partial M_{kb}}$.
Next, set $a\mapsto k$ in (\ref{eom}) and apply
$\frac{N(E_b+E_k)}{\mathrm{i}} \frac{\partial}{\partial M_{ak}}$.  Take
the difference of both equations and sum over $k$:
\begin{align}
-N\sum_{k=1}^{N} 
(E_a-E_b)\frac{\partial^2 
\mathcal{Z}(M)}{\partial 
M_{ak} \partial M_{kb}} 
=  \sum_{k=1}^{N} 
\Big(
M_{ka} \frac{\partial\mathcal{Z}(M)}{\partial M_{kb}} 
-M_{bk} \frac{\partial\mathcal{Z}(M)}{\partial M_{ak}} 
\Big)\;.
\label{WTI}
\end{align}
This is a Ward-Takahashi identity first discovered in \cite{Disertori:2006nq}.
The strategy which we follow here was suggested in \cite{Grosse:2012uv}. 
We insert (\ref{calZ4}) into (\ref{WTI}) and evaluate the derivatives 
for $a\neq b$:
\begin{align}
&\frac{1}{N}\sum_{k=1}^{N} 
\Big\{
(G_{|kb|} - G_{|ak|}) M_{bk} M_{ka} 
+ \frac{1}{N}(G_{|b|k|} -G_{|a|k|}) M_{ba} M_{kk}  \Big\}
\nonumber
\\
&=\frac{1}{N}\sum_{k=1}^{N} (E_a-E_b)
\Big\{ \frac{1}{N} \sum_{l=1}^{N} G_{|bkal|} 
+G_{|ak|} G_{|bk|} 
\nonumber
\\[-0.5ex]
&\qquad\qquad +\frac{1}{N^2} (G_{|b|abk|} +
G_{|a|abk|} + G_{|bk|ak|} 
\Big\}
M_{bk} M_{ka}
\nonumber
\\
&+ \frac{1}{N^2}\sum_{k=1}^{N} (E_a-E_b) 
\Big\{G_{|bakk|} 
+ \frac{1}{N}\sum_{l=1}^{N} G_{|k|bal|} 
+ G_{|ab|} (G_{|b|k|}+ G_{|a|k|}) 
\nonumber
\\[-0.5ex]
&\qquad\qquad + \frac{1}{N^2} (G_{|b|k|ab|}+G_{|a|k|ab|})
\Big\}
M_{ba} M_{kk}+\mathcal{O}(M^4)\;.
\label{WTI-2}
\end{align}

For the next steps we assume that the functions $G_{..k_i..}$ under
consideration are evaluations of holomorphic functions in several
complex variables at points $E_{k_i}$ in the holomorphicity
domain. See the discussion after (\ref{2pt}).  Applying to
(\ref{WTI-2}) the operators
$N \frac{\partial^2}{\partial M_{bp} \partial M_{pa}}$ or
$N^2\frac{\partial^2}{\partial M_{ba} \partial M_{pp}}$ for
$a\neq p\neq b$ gives two independent equations. Under the
holomorphicity assumption they extend continuously to $p=a$ and
$p=b$. After exchanging $p\leftrightarrow b$, these equations read
\begin{subequations}
\begin{align}
-\frac{G_{|pb|} - G_{|ab|}}{E_p-E_a}
&=
\frac{1}{N} \sum_{k=1}^{N} G_{|bakp|} 
+G_{|ab|} G_{|bp|} 
+\frac{1}{N^2} (G_{|p|bap|} +
G_{|a|bap|} + G_{|bp|ab|}) \;, \raisetag{1ex}
\label{WI2}
\\[-1.5ex]
-\frac{G_{|p|b|} - G_{|a|b|}}{E_p-E_a}
&=
G_{|bbpa|} 
+ \frac{1}{N}\sum_{k=1}^{N} G_{|b|akp|} 
+ G_{|ap|} (G_{|p|b|}+ G_{|a|b|}) 
\nonumber
\\*[-0.5ex]
&\qquad+ \frac{1}{N^2} (G_{|b|p|ap|}+G_{|a|b|ap|})\;;
\label{WI11}
\end{align}
\end{subequations}
they hold for $p\neq a$. By the holomorphicity assumption
the equations (\ref{WI2}) and (\ref{WI11}) extend continuously 
to $p=a$. Then, summing (\ref{WI2}) over $p$ collects 
the double-underlined terms in (\ref{DS2}) into
$-\frac{1}{N}\sum_{p=1}^{N} 
\frac{(G_{|pb|} - G_{|ab|})}{(E_p-E_a)}$, and the case $p=b$ of  
(\ref{WI11}) collects the single-underlined terms in 
 (\ref{DS2}). Similarly,  summing (\ref{WI11}) over $p$ collects 
the single-underlined terms in (\ref{DS11}) into
$-\frac{1}{N}\sum_{p=1}^{N} 
\frac{(G_{|p|b|} - G_{|a|b|})}{(E_p-E_a)}$, and the case $p=b$ of  
(\ref{WI2}) collects the double-underlined terms in 
 (\ref{DS11}):
\begin{align}
(E_a+E_b)G_{|ab|}&=1
-\frac{\lambda}{N} 
\sum_{p=1}^{N} G_{|ab|}G_{|ap|}
+\frac{\lambda}{N} 
\sum_{p=1}^{N}\frac{G_{|pb|}-G_{|ab|}}{E_p-E_a}
\nonumber
\\*[-0.5ex]
& - 
\frac{\lambda}{N^2}
\Big(-\frac{G_{|b|b|}-G_{|a|b|}}{E_b-E_a}+
G_{|abab|} +G_{|baaa|} +G_{|ab|}G_{|a|a|}
\nonumber
\\*[-0.5ex]
&\qquad +\frac{1}{N} \sum_{p=1}^{N} 
G_{|ab|ap|}\Big)
-\frac{\lambda}{N^4} G_{|a|a|ab|}\;,
\label{DS2-final}
\\
(E_a+E_a)G_{|a|b|}&=
-\frac{\lambda}{N} 
\sum_{p=1}^{N} G_{|ap|}G_{|a|b|}
+\frac{\lambda}{N} 
\sum_{p=1}^{N}\frac{G_{|p|b|}-G_{|a|b|}}{E_p-E_a}
+\lambda
\frac{G_{|bb|}-G_{|ab|}}{E_b-E_a}
\nonumber
\\[-1ex]
& 
-\frac{\lambda }{N^2} 
\Big(G_{|b|aaa|}+G_{|a|abb|}
+3 G_{|a|b|}G_{|a|a|}
+\frac{1}{N}
\sum_{p=1}^{N} G_{|a|b|ap|}\Big)
\nonumber
\\[-1ex]
&-\frac{\lambda}{N^4} G_{|b|a|a|a|}\;.
\label{DS11-final}
\end{align}
These identities have been found in \cite{Grosse:2012uv} (by a faster, 
but less elementary approach). 

Identities of such type
can be solved by a further expansion of all arising 
functions $G_{\dots}$ as formal power series in $N^{-2}$, 
\begin{align}
G_{\dots}=\sum_{g=0}^\infty \frac{1}{N^{2g}} G^{(g)}_{\dots}\;.
\end{align}
With the convention that $\frac{1}{N}
\sum_1^{N}$ is of order $N^0$, the coefficient of $N^{-2g}$ in
(\ref{DS2-final}) reads
\begin{align}
(E_a+E_b)G_{|ab|}^{(g)}&=\delta_{g,0}
-\frac{\lambda}{N} 
\sum_{p=1}^{N} \sum_{g_1+g_2=g} G_{|ab|}^{(g_1)}G_{|ap|}^{(g_2)}
+\frac{\lambda}{N} 
\sum_{p=1}^{N}\frac{G_{|pb|}^{(g)}-G_{|ab|}^{(g)}}{E_p-E_a}
\nonumber
\\*[-0.5ex]
& - 
\lambda
\Big(-\frac{G_{|b|b|}^{(g-1)}-G_{|a|b|}^{(g-1)}}{E_b-E_a}+
G_{|abab|}^{(g-1)} +G_{|baaa|}^{(g-1)}
+\sum_{g_+g_2=g-1} G_{|ab|}^{(g_1)}G_{|a|a|}^{(g_2)}
\nonumber
\\*[-0.5ex]
&\qquad +\frac{1}{N} \sum_{p=1}^{N} 
G_{|ab|ap|}^{(g-1)}\Big)
-\lambda G_{|a|a|ab|}^{(g-2)}\;.
\label{DS2-final-g}
\end{align}
For the degree or genus $g=0$ we thus obtain the closed 
equation (\ref{2pt})  for $G_{|ab|}^ {(0)}$. Similarly,  the restriction of 
(\ref{DS11-final}) to the degree or genus $g=0$ is (\ref{G11}). 
Both equations have been solved in this paper.

%\bibliographystyle{halpha-abbrv}
%\bibliography{integrability}

\end{document}